\documentclass[a4paper,11pt]{article}
\usepackage[utf8]{inputenc}
\usepackage{color}
\usepackage{amssymb}
\usepackage{amsmath}
\usepackage{graphicx}
\usepackage{mathtools}
\usepackage{ragged2e}
\usepackage{hyperref}
\hypersetup{
    colorlinks=true,
    linkcolor=blue,
    filecolor=cyan, 
    urlcolor=red,
    citecolor=red,
} 
\usepackage[titletoc,toc,title]{appendix}
\numberwithin{equation}{section}
\usepackage{cleveref}
\usepackage[margin=2.5cm]{geometry}
\usepackage{cite}
\usepackage{epsfig}
\usepackage{float}
\usepackage{cancel}
\usepackage{amsfonts}
\usepackage{enumitem}
\usepackage[font={footnotesize,it}]{caption}
\usepackage{authblk}

\begin{document}
\begin{titlepage}
\title{Holographic Entanglement Negativity for Disjoint Subsystems in Conformal Field Theories with a Conserved Charge}
\date{}

\author[1]{Mir Afrasiar\thanks{\noindent E-mail:~ afrasiar@iitk.ac.in}}
\author[1]{Jaydeep Kumar Basak\thanks{\noindent E-mail:~ jaydeep@iitk.ac.in}} 
\author[1]{Vinayak Raj\thanks{\noindent E-mail:~ vraj@iitk.ac.in}}
\author[1]{Gautam Sengupta\thanks{\noindent E-mail:~ sengupta@iitk.ac.in}}

\affil[1]{
Department of Physics\\

Indian Institute of Technology\\ 

Kanpur, 208016\\ 

India}

\maketitle
\begin{abstract}
\noindent
\justify

We investigate the extension of a holographic construction for the entanglement negativity of two disjoint subsystems in proximity to $CFT_d$s with a conserved charge dual to bulk $AdS_{d+1}$ geometries. The construction involves a specific algebraic sum of the areas of bulk co-dimension two static minimal surfaces  homologous to certain appropriate combinations of the subsystems in question. In this connection we compute the holographic entanglement negativity for two disjoint subsystems in proximity, with long rectangular strip geometries in $CFT_d$s dual to bulk non extremal and extremal RN-$AdS_{d+1}$ black holes. Our results conform to quantum information theory expectations and also reproduces earlier results for adjacent subsystems in the appropriate limit which constitutes strong consistency checks for our holographic construction.

\end{abstract}
\end{titlepage}
\tableofcontents
\pagebreak

\section{Introduction}
\label{sec1}
\justify

Over last two decades quantum entanglement has emerged as a central theme in widely ranging areas of physics from strongly correlated many body condensed matter systems to the black hole information loss paradox. The entanglement for bipartite pure states is characterized by the entanglement entropy which is defined in quantum information theory as the von Neumann  entropy of the reduced density matrix for the subsystem under consideration. Although the computation of this quantity is straightforward for finite quantum systems, the corresponding issue for quantum many body systems involve infinite number of eigenvalues for the reduced density matrix and is hence intractable. For such cases a formal definition of the entanglement entropy may be possible through an appropriate {\it replica technique} although its explicit evaluation is difficult in general. Interestingly for conformally invariant 
$(1+1)-$dimensional quantum field theories ($CFT_{1+1}$) it was possible to explicitly compute the entanglement entropy for bipartite states through a suitable replica technique developed in  \cite{Calabrese:2004eu, Calabrese:2009qy}. 

For bipartite mixed states the entanglement entropy receives contributions from irrelevant correlations and is hence unable to correctly characterize the entanglement for such states. This issue was addressed in quantum information theory and several entanglement measures such as \textit{distillable entanglement, entanglement of purification} etc. were proposed to correctly characterize mixed state entanglement \cite{Plenio:2007zz} but were difficult to compute as they were given by variational expressions over LOCC protocols. Interestingly a computable measure for mixed state entanglement was proposed by Vidal and Werner in a seminal work \cite{PhysRevA.65.032314} and this novel measure was termed as {\it entanglement negativity} which was given by the logarithm of the trace norm for the partially transposed reduced density matrix with respect to one of the subsystems of the bipartite system in question. This quantity provided an upper bound on the distillable entanglement and it was shown by Plenio in \cite{Plenio:2005cwa} that it was non convex and an entanglement monotone under LOCC. Interestingly in \cite{Calabrese:2012nk,Calabrese:2012ew,Calabrese:2014yza} the authors obtained the entanglement negativity for various bipartite pure and mixed states in $CFT_{1+1}$s using a modified version of  the replica technique for the entanglement entropy. They showed that for mixed state configurations of two or more disjoint intervals in a zero temperature $CFT_{1+1}$ and a single interval in a finite temperature $CFT_{1+1}$ the entanglement negativity involved a non-universal contribution which depends on the full operator content of the theory. However, in the large central charge limit these non universal contributions were sub leading and it was possible to extract a universal contribution to the entanglement negativity for such mixed states in $CFT_{1+1}$s.
 
The explicit computation of entanglement measures through replica techniques for $CFT_{1+1}$ described above inspired a holographic characterization of these results in terms of the bulk geometry in the context of the $AdS/CFT$ correspondence. In this connection Ryu and Takayanagi (RT) \cite{Ryu:2006bv, Ryu:2006ef} in a classic work advanced a holographic entanglement entropy conjecture which described the universal part of the entanglement entropy for a subsystem in a $CFT_d$ to be proportional to the area of the co-dimension two bulk static minimal surface homologous to the subsystem, in the dual $AdS_{d+1}$ geometry.  A covariant generalization of the RT conjecture was subsequently proposed in \cite{Hubeny:2007xt} by Hubeny, Rangamani and Takayanagi (HRT) which characterized the holographic entanglement entropy in terms of the corresponding co-dimension two bulk extremal surface homologous to the subsystem in question. Proofs of the RT and the HRT conjectures were developed in a series of articles described in \cite{Fursaev:2006ih, Headrick:2010zt, Casini:2011kv, Faulkner:2013yia, Lewkowycz:2013nqa, Hartman:2013mia, Dong:2016hjy}. These fascinating developments inspired intense investigations in this field which led to significant insights into the structure of entanglement for bipartite pure states in holographic $CFT_d$s \cite {Nishioka:2009un, Cadoni:2010kla, Fischler:2012ca, Fischler:2012uv, Takayanagi:2012kg, Chaturvedi:2016kbk}.

As mentioned earlier the entanglement entropy was not a viable measure for the characterization of mixed state entanglement. In the context of the exciting developments regarding the holographic description of the entanglement entropy described above it was natural to seek a corresponding holographic characterization of the entanglement negativity in dual $CFT_d$s which described mixed state entanglement in quantum information theory. This important question was investigated in \cite{Rangamani:2014ywa} where the holographic entanglement negativity for the bipartite pure vacuum state of a $CFT_d$ could be obtained although a general holographic prescription for mixed states in $CFT_d$ remained an elusive issue. This significant issue was addressed in \cite{Chaturvedi:2016rcn} where the authors (CMS) proposed a holographic characterization of the entanglement negativity for bipartite states described by a single subsystem in $ CFT_d$s dual to bulk $AdS_{d+1}$ geometries. In the context of the $AdS_3/CFT_2$ scenario their conjecture involved a specific algebraic sum of the lengths of bulk space like geodesics homologous to certain combinations of intervals. The application of their proposal exactly reproduced the corresponding replica technique results in the large central charge limit for holographic $CFT_{1+1}$s. A higher dimensional generalization of their proposal involved the corresponding algebraic sum of the areas of bulk co dimension two static minimal surfaces homologous to the combinations of such subsystems \cite{Chaturvedi:2016rft} which reproduced universal features of entanglement negativity observed for the corresponding lower dimensional holographic $CFT_{1+1}$s and were in conformity with quantum information theory expectations. A covariant generalization of the proposal was later developed  in \cite{Chaturvedi:2016opa} in terms of the areas of such bulk codimension two extremal surfaces. The holographic proposal in \cite{Chaturvedi:2016rcn} was further substantiated in \cite{Malvimat:2017yaj} through a large central charge analysis of the $CFT_{1+1}$ results utilizing the monodromy technique \cite{Hartman:2013mia, Fitzpatrick:2014vua}. 

Subsequent to the developments discussed above the proposal for the holographic entanglement negativity for dual $CFT_d$s and its covariant generalization was further extended to bipartite mixed state configuration of adjacent interval in dual $CFT_{1+1}$\cite{Jain:2017aqk,Jain:2017uhe}. This was later generalized  in \cite{Jain:2017xsu,Jain:2018bai} for the holographic entanglement negativity of two adjacent subsystems with long rectangular strip geometry in holographic $CFT_d$s at zero temperature, finite temperature and also in $CFT_d$s with a conserved charge dual to bulk pure $AdS_{d+1}$ geometries, $AdS_{d+1}$-Schwarzschild black holes and $AdS_{d+1}$-Reissner-Nordstr\"om (RN) black holes respectively. In a similar fashion the corresponding holographic entanglement negativity for a mixed state configuration of two disjoint intervals in proximity in $CFT_{1+1}$s dual to static bulk $AdS_3$ geometries was proposed in\cite{Malvimat:2018txq}. A covariant generalization of their proposal  was described in \cite{Malvimat:2018ood} and was subsequently generalized to higher dimensions in \cite{Basak:2020bot} for subsystems with such long rectangular strip geometries in  $CFT_d$s dual to bulk pure $AdS_{d+1}$ geometries and $AdS_{d+1}$-Schwarzschild black holes. 
Interestingly for all these examples the holographic entanglement negativity obtained through the application of the above constructions reproduced universal features upto the leading order observed for lower dimensions which were also in agreement with quantum information theory expectations. We mention here that very interestingly the holographic entanglement negativity conjecture described above have also been recently generalized in the context of flat space holography in \cite{Basu:2021axf} for $(1+1)$-dimensional non relativistic Galilean conformal field theories dual to $(2+1)$-dimensional bulk asymptotically flat geometries and their results exactly reproduce the corresponding field theory replica technique computations described in  \cite{Malvimat:2018izs} in the large central charge limit. Furthermore a strong substantiation of their construction in \cite{Basu:2021axf} was also established through a comprehensive large central charge analysis involving the geometric monodromy construction. 

As mentioned earlier a rigorous consistency check for the holographic entanglement negativity construction was established through a large central charge analysis for the $AdS_3/CFT_2$ scenario in \cite{Malvimat:2017yaj} although a proof for generic $AdS_{d+1}/CFT_d$ scenario along the lines of \cite{Faulkner:2013yia, Lewkowycz:2013nqa, Dong:2016fnf} from a bulk perspective was a non trivial open problem. Very recently this significant issue has been partially addressed in the context of a bulk gravitational path integral in \cite{Basak:2020aaa}  involving certain replica symmetry breaking saddles  described in \cite{Dong:2021clv} for spherical entangling surfaces. A general bulk proof for the holographic entanglement negativity conjecture for spherical entangling surfaces may be deduced from the arguments presented in \cite{Basak:2020aaa}. However an extension of this proof for arbitrary subsystem geometries is still an involved issue.

In this article, we extend the conjecture described in \cite{Malvimat:2018txq, Basak:2020bot} for various mixed state configurations of disjoint subsystems in proximity with long rectangular strip geometries in holographic $CFT_d$s with a conserved charge dual to bulk non-extremal and extremal RN-$AdS_{d+1}$ black holes. Our computations involve a perturbative expansion of the areas of co dimension two static minimal surfaces for certain non trivial limits of suitable thermodynamic parameters describing the bulk configuration. In order to define the structure of this non trivial perturbative expansion we first consider the simpler $AdS_4/CFT_3$ scenario and subsequently address the issue of the more general $AdS_{d+1}/CFT_d$ framework. The results for both the scenarios considered here demonstrate a consistent uniform behaviour and also conforms to the expectations from quantum information theory for the entanglement negativity which serves as an important consistency check. Furthermore our results match with those for adjacent subsystems described in \cite{Jain:2018bai} in the appropriate limit either exactly or up to the leading order. Note that there is a subtle difference with the results obtained in \cite{Jain:2018bai} where the perturbative expansion involves the $\mathcal{O}(\epsilon^0)$ term along with other $\mathcal{O}(\epsilon)$ terms where $\epsilon$ is a generic small parameter. However the corresponding perturbative expansion for disjoint subsystems begins with the $\mathcal{O}(\epsilon)$ terms. The reason for this important difference is that adjacent subsystems involve a shared boundary which serves as the entangling surface and incorporates a large number of entangled degrees of freedom which provides a dominant contribution to the entanglement negativity and is manifested in the $\mathcal{O}(\epsilon^0)$ term with the higher order terms serving as small corrections. However, disjoint subsystems do not incorporate such a shared boundary which results in the perturbative expansion involving only higher order terms. Interestingly our results for the holographic entanglement negativity correctly reproduces this scenario and also serves as a strong substantiation for our construction.

On a separate note, the authors in \cite{Kudler-Flam:2018qjo} proposed the backreacted minimal entanglement wedge cross section (EWCS)  as the holographic dual of the entanglement negativity in the $AdS_{d+1}/CFT_d$ scenario for subsystems with spherical symmetry.\footnote{Note that the minimal EWCS has also been proposed as the holographic dual to the odd entanglement entropy \cite{Tamaoka:2018ned, Kusuki:2019rbk, Kusuki:2019evw, Mollabashi:2020ifv}, balanced partial entanglement \cite{Wen:2021qgx} and entanglement of purification in \cite{Nguyen:2017yqw, Takayanagi:2017knl}.} Subsequently, a derivation of their conjecture was also advanced in \cite{Kusuki:2019zsp} inspired by the idea of the {\it reflected entropy} described in \cite{Dutta:2019gen}.  A further qualification of their conjecture for the case of a single interval in an infinite system at a finite temperature in the $AdS_3/CFT_2$ scenario
was considered in \cite{Basak:2020oaf} where a resolution for a mismatch with the replica technique results in \cite{Calabrese:2014yza} was provided through an alternate construction for the EWCS. It was shown in \cite{Basak:2020oaf} that the two proposals were completely equivalent for the $AdS_3/CFT_2$ scenario.\footnote{See also \cite{BabaeiVelni:2019pkw,BabaeiVelni:2020wfl,Sahraei:2021wqn} for other developments regarding EWCS in higher dimensions.}

Furthermore in the recent past, the authors in \cite{KumarBasak:2021lwm} have proposed a covariant generalization of the holographic entanglement negativity conjecture based on the EWCS and their results also exactly reproduced the results for the holographic entanglement negativity obtained through the earlier construction further demonstrating the equivalence of the two proposals for the $AdS_3/CFT_2$ scenario. It could also be inferred from their analysis that for higher dimensions
the only difference between the two proposals was in an overall dimension dependent numerical constant arising from the backreaction of the cosmic brane for the bulk conical defect geometry in the replica limit.  However this backreaction factor could only be determined explicitly for subsystems with spherical symmetry as described in \cite{Kudler-Flam:2018qjo, Kusuki:2019zsp}.
Consequently our results in the present article are also expected to be modified by such an overall dimension dependent constant  backreaction factor. However the explicit determination of this factor for the subsystem geometries of long rectangular strips considered by us remains a non trivial open issue. Note that this overall constant factor does not change the corresponding physics.

This article is organized as follows, in section \ref{sec2} we describe the holographic entanglement negativity construction for mixed state configurations of disjoint subsystems. The holographic entanglement negativity for mixed states of two disjoint subsystems in proximity for the $AdS_4/CFT_3$ case is computed in section \ref{sec3}. Subsequently, in section \ref{sec4} we utilize our construction to obtain the holographic entanglement negativity for such mixed states in the general $AdS_{d+1}/CFT_d$ scenario. Finally, in section \ref{sec5} we summarize our results and present our discussions.

\section{Holographic entanglement negativity }\label{sec2}

We begin by briefly reviewing the holographic entanglement negativity construction for bipartite mixed states of disjoint subsystems in $CFT_d$s dual to bulk static $AdS_{d+1}$ geometries as described in \cite{Basak:2020bot}. To this end we consider a tripartite system $ A \cup B$ in a pure state in the dual $CFT_d$ where $A=A_1\cup A_2$ is a disjoint bipartite subsystem and $B$ describes the rest of the system. We denote $A_s$ ($A_s\subset B$) as the subsystem separating $A_1$ and $A_2$. To illustrate the holographic construction we first consider the above configuration in the context of the $AdS_3/CFT_2$ scenario where the subsystems are essentially intervals.

For the mixed state configuration of $A=A_1\cup A_2$ described by the disjoint intervals $A_1,A_2$ the 
entanglement negativity is defined as 
\begin{align}
\mathcal{E}=ln~Tr|\rho_A^{T_2}|,
\end{align}
where, $\rho_A^{T_2}$ is the partial transpose of the reduced density matrix $\rho_A$ with respect to the interval $A_2$.
We can explicitly determine $Tr(\rho_A^{T_2})^{n_e}$ through a replica technique \cite{Calabrese:2012nk,Calabrese:2012ew}, where $n_e$ denotes the even parity of the replica index $n$. Subsequently we can obtain the entanglement negativity for the subsystem $A$ through an analytic continuation of the even sequence of $n$ to $n_e \to 1$ as
\begin{align}
\mathcal{E}=\lim_{n_e\to 1}~ln~Tr(\rho_A^{T_2})^{n_e}.
\end{align}
For the above configuration the quantity $Tr(\rho_A^{T_2})^{n_e}$ is given by a four point correlation function of certain twist fields as,
\begin{align}
{Tr}(\rho_A^{T_2})^{n_e} = 
\langle\mathcal{T}_{n_e}(z_1)\overline{\mathcal{T}}_{n_e}(z_2)\overline{\mathcal{T}}_{n_e}(z_3)\mathcal{T}_{n_e}(z_4)\rangle_{\mathbb{C}},
\end{align}
where $\mathbb{C}$ denotes the complex plane. The twist fields $\mathcal{T}, \overline{\mathcal{T}}$s essentially
imposes the boundary conditions between the $n_e$ replica and reduce to primary fields on the complex plane in the replica limit. In the large central charge limit, this twist correlator in the $t-channel$ can be expressed as 
\begin{equation}\label{large_c4point}
\lim_{n_e\to 1}\left\langle{\mathcal T}_{n_e}(z_1){\overline{\mathcal T}}_{n_e}(z_2){\overline{\mathcal T}}_{n_e}(z_3){\mathcal T}_{n_e}(z_4)\right\rangle_{ \mathbb{C}}=\left(1-x\right)^{2\hat h},
\end{equation}
where, $x=\frac{|z_1-z_2||z_3-z_4|}{|z_1-z_3||z_2-z_4|}$ is the cross ratio, $\hat h$ is the conformal dimension and the two disjoint intervals are considered to be in proximity ($\frac{1}{2}<x<1$). The two point correlator of the twist fields in the $CFT_2$ is given as
\begin{align}\label{2point_fn}
\left\langle{\mathcal T}_{n_e}(z_i)\overline{{\mathcal T}}_{n_e}(z_j)\right\rangle_{\mathbb{C}}\sim|z_{ij}|^{-2\Delta_{{\mathcal T}_{n_e}}} ,
\end{align}
where, $z_{ij}=z_i-z_j$ and $\Delta_{\mathcal 
	{T}_{n_e}}$ is the scaling dimension of the twist field $\mathcal{T}_{n_e}$. Using the $AdS/CFT$ dictionary 
this two point correlator may be expressed in terms of the lengths of spacelike geodesics $\mathcal{L}_{ij}$ in the 
bulk $AdS_3$(with length scale $R$) geometry as
\begin{align}\label{2point-fn-bulk}
\left\langle{\mathcal T}_{n_e}(z_i)\overline{{\mathcal T}}_{n_e}(z_j)\right\rangle_{\mathbb{C}}\sim\exp\left(-\frac{\Delta_{{\mathcal T}_{n_e}}{\mathcal L}_{ij}}{R}\right).
\end{align}
Employing eqs. (\ref{2point_fn}) and (\ref{2point-fn-bulk}), the four point correlator (\ref{large_c4point}) may thus be written as
\begin{align}\label{HEN}
\lim_{n_e\to 1}\left\langle{\mathcal T}_{n_e}(z_1)\overline{{\mathcal T}}_{n_e}(z_2)\overline{{\mathcal T}}_{n_e}(z_3){\mathcal T}_{n_e}(z_4)\right\rangle_{\mathbb{C}} 
= \exp\left[\frac{ c }{ 8 R }\left({\mathcal L}_{13}+{\mathcal L}_{24}-{\mathcal L}_{14}-{\mathcal L}_{23}\right)\right].
\end{align}
Subsequently utilizing the Brown-Henneaux formula $c=\frac{3R}{2G_N^3}$, the holographic entanglement negativity for the mixed state of the disjoint intervals in proximity is given by
\begin{align}\label{HEN2}
{\mathcal E} &= \frac{3}{16 G_N^{(3)}}\left({\mathcal L}_{A_1\cup A_s}+{\mathcal L}_{A_2\cup A_s}-{\mathcal L}_{A_1\cup A_2\cup A_s}-{\mathcal L}_{A_s}\right)
\end{align}

The above holographic  construction may be generalized to the higher dimensional $AdS_{d+1}$/ $CFT_d$ framework
in terms of the areas of bulk codimension two static minimal surfaces and the corresponding holographic entanglement negativity for the above configuration may be expressed as follows
\begin{align}\label{hen2}
\mathcal{E}=\frac{3}{16G_N^{(d+1)}}\bigg( \mathcal{A}_{A_1\cup A_s}+\mathcal{A}_{A_s\cup A_2}-\mathcal{A}_{A_1\cup A_s\cup A_2}-\mathcal{A}_{A_s} \bigg).
\end{align}
Here, $G_N^{(d+1)}$ is the gravitational constant in ($d+1$)-dimension and $A_\gamma $ denotes the area of a bulk codimension two static minimal surface homologous to any subsystem $\gamma$ described in \cite{Basak:2020bot}. Using the Ryu Takayanagi conjecture $\big(\mathcal{S}_{A_i}=\frac{\mathcal{A}_i}{4G_N^{(d+1)}}\big)$ we can rewrite the eq.(\ref{hen2}) as follows,
\begin{equation}\label{hen3}
\mathcal{E}=\frac{3}{4}\bigg( \mathcal{S}_{A_1\cup A_s}+\mathcal{S}_{A_s\cup A_2}-\mathcal{S}_{A_1\cup A_s\cup A_2}-\mathcal{S}_{A_s} \bigg).
\end{equation}
Note that the holographic entanglement negativity in eq.(\ref{hen3}) can be expressed as,
\begin{equation}
\mathcal{E}=\frac{3}{4}\bigg( \mathcal{I}(A_1\cup A_s,A_2)-\mathcal{I}(A_s,A_2) \bigg). 
\end{equation}
where $\mathcal{I}(A_i,A_j)=\mathcal{S}_{A_i}+\mathcal{S}_{A_j}-\mathcal{S}_{A_i\cup A_j}$ is
the holographic mutual information between two subsystems $A_i$ and $A_j$. In the rest of the work we proceed to apply the above construction to obtain the holographic entanglement negativity for disjoint subsystems in proximity in $CFT_d$s with a conserved charge dual to bulk $AdS_{d+1}$ geometries.

\section{Holographic entanglement negativity for $\mathrm{CFT_3}$ with a conserved charge dual to RN-$\mathrm{AdS_4}$}\label{sec3}

As an exercise to set the notations and the formulation of the issue for the generic $AdS_{d+1}/CFT_d$ scenario,
we first study the  application of the holographic construction described above to obtain the entanglement negativity for the bipartite mixed state configuration of disjoint subsystems in proximity, with long rectangular strip geometries
in $CFT_3$s with a conserved charge dual to bulk RN-$AdS_4$ black holes. In this context we provide a detailed analysis
of the perturbative expansion for the areas of the corresponding bulk codimension two static minimal surfaces, for
different regimes of the charge and the temperature of the dual $CFT_3$. 

\subsection{Minimal surface area for RN-$\mathrm{AdS_4}$ black holes}

In this subsection we obtain an expression for the area of a bulk codimension two static minimal surface  homologous to a subsystem with long rectangular strip geometry in the $CFT_3$ with a conserved charge dual to bulk RN-$AdS_4$ black hole. Setting the $AdS$ radius $R=1$, RN-$AdS_4$ metric is given by,
\begin{eqnarray}
ds^2 &=& -r^2f(r)dt^2+\frac{1}{r^2f(r)} dr^2+{r^2}d\Omega^2,\label{RNmetric}\\
f(r)&=& 1-\frac{M}{r^3}+\frac{Q^2}{r^4}.\label{lapsefnc}
\end{eqnarray}
As the lapse function $f(r)$ vanishes at the horizon $r=r_h$ 
\begin{equation}\label{MQrel}
f(r_{h})=0 \Rightarrow M=\frac{r_{h}^4+Q^2}{r_{h}}.
\end{equation}
In terms of the charge $Q$ and the horizon radius $r_h$ we can rewrite the lapse function as follows,
\begin{equation}\label{RNlapse1}
f(r)=1-\frac{r_h^3}{r^3}-\frac{Q^2}{r^3r_h}+\frac{Q^2}{r^4}.
\end{equation}
Using the above expression, the Hawking temperature is defined as,
\begin{equation}\label{RNtemp}
T=\left.\frac{f'(r)}{4\pi}\right|_{r=r_h}=\frac{3 r_{h}}{4\pi}(1-\frac{Q^{2}}{3 r_{h}^{4}}).
\end{equation}
We now specify the subsystem $A$ with a rectangular strip geometry in the dual $CFT_3$ in terms of the coordinates as follows
\begin{equation}\label{subsys}
x\in[-\frac{l}{2},\frac{l}{2}],~~~~~y\in[-\frac{L}{2},\frac{L}{2}].
\end{equation}
Then the area ${\cal A}_A$ of the co-dimension two bulk static minimal surface, homologous to the subsystem $A$, is given as 
\begin{equation}\label{Earea1}
{\cal A}_A=2 L\int^\infty_{r_{c}} \frac{ dr}{\sqrt{f(r)(1-\frac{r_{c}^{4}}{r^{4}}})}.
\end{equation}
Note that the length $l$ of the rectangular strip in the $x$-direction and the turning point $r_{c}$ in the bulk for the co-dimension two static minimal surface are related to each other as follows
\begin{equation}\label{lrel}
\frac{l}{2}=\int^\infty_{r_{c}} \frac{r_{c}^{2} dr}{r^{4}\sqrt{f(r)(1-\frac{r_{c}^{4}}{r^{4}})}}.
\end{equation}

For later convenience we implement a coordinate transformation from $r$ to $u=\frac{r_c}{r}$ for evaluation of the above integrals. The eq. (\ref{RNlapse1}) and the integrals (\ref{Earea1}) and (\ref{lrel}) may then be expressed as follows
\begin{eqnarray}
f(u)&=& 1-\frac{{r_h}^3 u^3}{{r_{c}}^3}-\frac{Q^2 u^3}{{r_{c}}^3 {r_h}}+\frac{Q^2 u^4}{{r_{c}}^4},\label{RNlapse2}\\
{\cal A}&=&2Lr_{c}\int _0^1\frac{f(u)^{-\frac{1}{2}} }{u^2\sqrt{1-u^4}}du,\label{Ainu}\\
l&=&\frac{2}{r_{c}}\int _0^1\frac{u^2 f(u)^{-\frac{1}{2}}}{\sqrt{1-u^4}}du. \label{linu}
\end{eqnarray}
The above integrals may only be evaluated exactly for the $CFT_3$ at a zero temperature in its vacuum state dual to a bulk pure $AdS_3$ geometry. For the case of the bulk RN-$AdS_4$ geometries with a non zero charge $Q$ and temperature $T$, these integrals may be evaluated perturbatively for different regimes of the above parameters which we will describe in the later sections.

\begin{figure}[H]
	\centering
	\includegraphics[scale=.3]{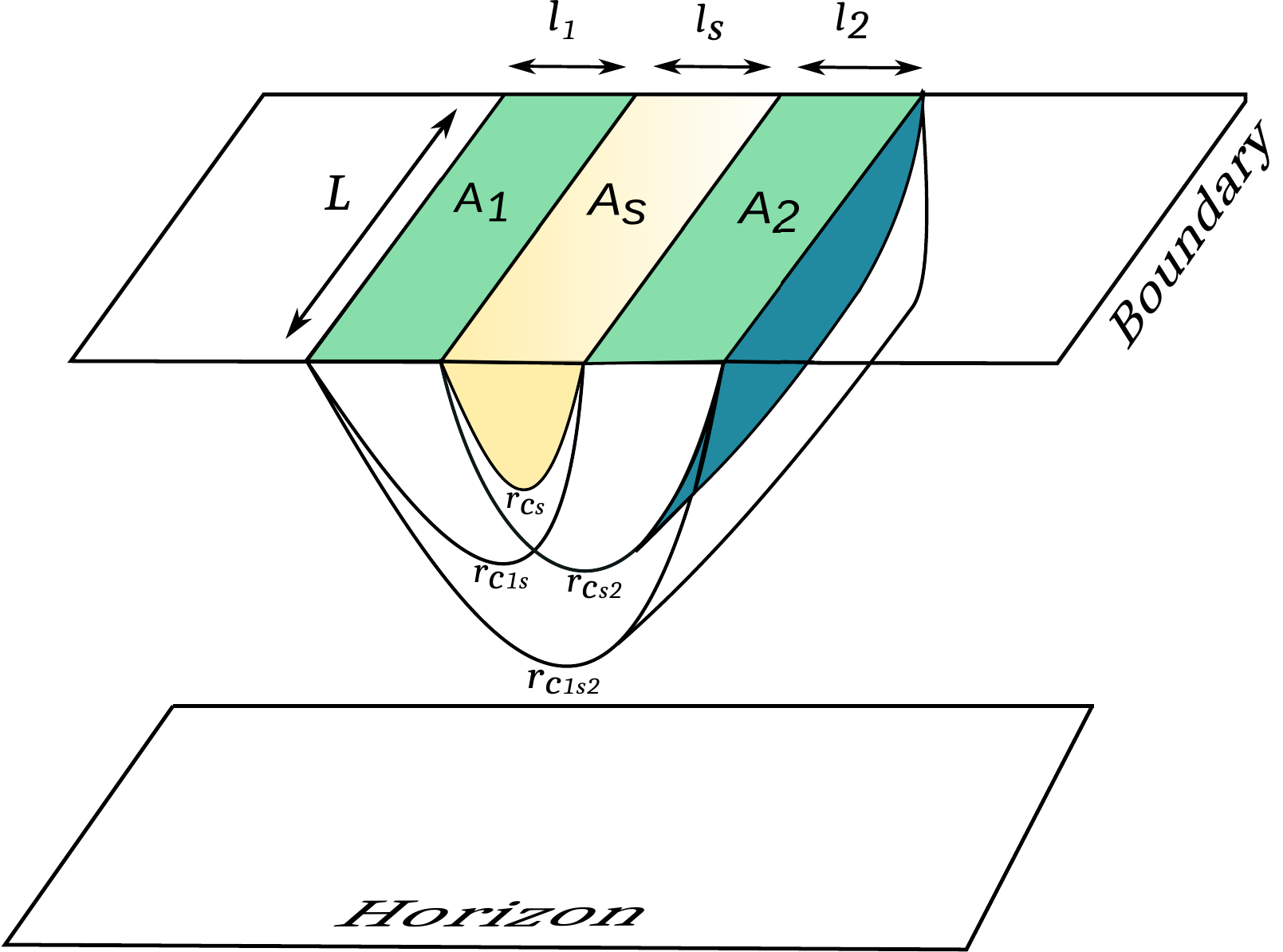}
	\caption{Schematic of the bulk static minimal surfaces that are homologous to the subsystems $A_1\cup A_s$, $A_s\cup A_2$, $A_1 \cup A_s \cup A_2$ and $A_s$ on the boundary $CFT_3$ dual to the RN-$AdS_4$ black hole. (Adapted from \cite{Basak:2020bot})}\label{fig1}
\end{figure}

Having obtained an expression for the area of a bulk co-dimension two static minimal surface with a long rectangular strip geometry we may now proceed to utilize the above to obtain the holographic entanglement negativity for disjoint subsystems through the eq. (\ref {hen2}).  The corresponding subsystems $A_1$, $A_s$ and $A_2$ are given as in
fig. (\ref{fig1}) as follows 
\begin{eqnarray}
&x\in[-\frac{l_1}{2},\frac{l_1}{2}],~~~~~y\in[-\frac{L}{2},\frac{L}{2}],\label{a1rec}\nonumber\\
&x\in[-\frac{l_s}{2},\frac{l_s}{2}],~~~~~y\in[-\frac{L}{2},\frac{L}{2}],\label{asrec}\\
&x\in[-\frac{l_2}{2},\frac{l_2}{2}],~~~~~y\in[-\frac{L}{2},\frac{L}{2}],\label{a2rec}\nonumber .
\end{eqnarray}
Now, from the equations (\ref{Ainu}) and (\ref{linu}), we can obtain the corresponding areas and the turning points of the co-dimension two bulk static minimal surfaces homologous to the subsystems $A_1$, $A_s$ and $A_2$ respectively by replacing $l$ in eq.(\ref{linu}) by $l_1$, $l_s$ and $l_2$.

\subsection{Non-extremal RN-$\mathrm{AdS_4}$ black holes}

We first consider the mixed state configuration of the disjoint subsystems in question, with long rectangular strip geometry
in $CFT_3$s with a conserved charge at a finite temperature, dual to bulk non-extremal RN-$AdS_4$ black holes as shown in fig. (\ref{fig1}). In order to obtain the holographic entanglement negativity for this configuration it is required to
evaluate the areas of the bulk co dimension two minimal surfaces, homologous to appropriate combinations of the subsystems in the dual $CFT_3$. For this purpose we utilize a perturbative technique to evaluate the corresponding areas of the RT surfaces for various regimes of  the charge $Q$ and the temperature $T$. 

\subsubsection{Small charge and low temperature}\label{small_Q_low_T}

For the non extremal black holes the non zero temperature $T$ leads to the following non-extremality condition from eq.(\ref{RNtemp}) as,
\begin{equation}\label{rhforlowtemp}
r_{h}>\frac{\sqrt{Q}}{{3^{\frac{1}{4}}}}. 
\end{equation}
In the small charge and low temperature limit we have the conditions $ r_{h} \ll r_{c} $ and $Q/r_{h}^2\sim 1$ which indicates that the RT surfaces are located far away from the black hole horizon \cite{Kundu:2016dyk}. In this case we can expand  $f(u)^{-\frac{1}{2}}$ where $f(u)$ is the lapse function, around $\frac{r_h}{r_c}=0$ to the leading order in $\mathcal{O}[(\frac{r_h}{r_c}u)^3]$ as follows\cite{Chaturvedi:2016kbk}
\begin{equation}\label{lapseRNads1}
f(u)^{-\frac{1}{2}}\approx1+\frac{1+\alpha}{2}\left(\frac{r_h}{r_c}\right)^3 u^3,
\end{equation}
where $\alpha=\frac{Q^2}{r_h^4}$.

The area of the bulk co-dimension two minimal surface homologous to the subsystem $A$ in the dual $CFT_3$ may then be expressed using the eqs. (\ref{lapseRNads1}), (\ref{linu}) and (\ref{Ainu}) as, \cite{Chaturvedi:2016kbk}
\begin{equation} \label{Area 3.2.1}
{\cal A}_A={\cal A}_A^{div}+{\cal A}_{ A}^{finite},
\end{equation}
where, ${\cal A}_A^{div}$ is the divergent part and ${\cal A}_{ A}^{finite}$ is the finite part of the area ${\cal A}_A$. 
These may be expressed as follows
\begin{equation} \label{A_div}
{\cal A}_A^{div}=2\Big(\frac{L}{a}\Big),
\end{equation}
\begin{equation} \label{A_finite}
{\cal A}_{ A}^{finite}= h_1\frac{L}{l}+h_2r_h^3(1+\alpha)l^2+\mathcal{O}(r_h^4 l^3),
\end{equation}
where $a$ is the UV cut-off of the boundary $CFT_3$ and the constants $h_1, h_2$ are given as, 
\begin{eqnarray}
h_1 &=& -\frac{ 4 \pi \Gamma(\frac{3}{4})^2}{\Gamma(\frac{1}{4})^2},\label{h1}\\
h_2 &=&  \frac{\Gamma (\frac{1}{4})^2 }{32\Gamma (\frac{3}{4})^2}\label{h2}.
\end{eqnarray}
We may now obtain the holographic entanglement negativity for the mixed state of the disjoint subsystems in proximity
utilizing the above expression for the area of the RT surface in eq. (\ref{hen2}) describing our proposal as follows
\begin{equation}\label{ENforSQlowtemp}
\mathcal{E}_\mathrm{disjoint} = \frac{3}{16G_{N}^{3+1}}\Big[h_1\Big(\frac{L}{l_1+l_s}+\frac{L}{l_s+l_2}-\frac{L}{l_1+l_s+l_2}-\frac{L}{l_s}\Big)-2h_2 M L ~l_1 l_2\Big] + \ldots,
\end{equation} 
where the first term arises from the vacuum configuration for the mixed state of disjoint subsystems in the corresponding $CFT_3$ with zero charge and at a zero temperature, dual to the bulk pure $AdS_4$ space time. The second term describes
the finite temperature sub leading contribution for a non zero conserved charge.

As a consistency check we may now consider the above expression in the limit of the subsystems being adjacent to each other with $l_s\rightarrow a$ where $(a \ll 1)$ is a UV cut off and setting $l_1+l_s\rightarrow l_1$, $l_s+l_2\rightarrow l_2$ and $l_1+l_s+l_2\rightarrow l_1+l_2$ we obtain,
\begin{equation}
\mathcal{E}_\mathrm{adjacent}=\frac{3}{16G_{N}^{3+1}}\Big[\Big(\frac{2L}{a}\Big) +h_1\Big(\frac{L}{l_1}+\frac{L}{l_2}-\frac{L}{l_1+l_2}\Big)-2h_2 M L ~l_1 l_2\Big] + \ldots.
\end{equation}
To obtain the first diverging term in the above expression we have used eqs. \eqref{Area 3.2.1}, \eqref{A_div} and \eqref{A_finite} in proper adjacent limit. This matches precisely with the holographic entanglement negativity for the mixed state of adjacent subsystems with rectangular strip geometries in a $CFT_3$ with a conserved charge, dual to bulk non-extremal RN-$AdS_4$ black hole as described in \cite{Jain:2018bai}.

\subsubsection{Small charge - high temperature}

Next we explore the small charge and the high temperature regime which leads to the following non extremality condition
$r_h l \gg 1$, and $\delta = \frac{Q}{\sqrt{3}r_h^2}\ll 1$. In this case we are required to Taylor expand the quantity
$f(u)^{-\frac{1}{2}}$  where $f(u)$ is the lapse function, around $\delta=0$ \cite{Chaturvedi:2016kbk} as follows
\begin{equation}\label{lapseRNads2}
f(u)^{-\frac{1}{2}}\approx \frac{1}{\sqrt{1-\frac{{r_h}^3 u^3}{{r_c}^3}}}+
\frac{3}{2}\left(\frac{r_h}{r_c}\right)^3\frac{\delta ^2  u^3 (1-\frac{{r_h} u}{{r_c}})}
{(1-\frac{{r_h}^3 u^3}{{r_c}^3})^{3/2})}. 
\end{equation}
Note that in the high temperature limit the RT surface approaches the black hole horizon in the bulk space time i.e. $r_h\sim r_c$ staying at a small finite distance above the horizon but never penetrates it \cite{Hubeny:2012ry,Kundu:2016dyk}.
This allows us to implement a near horizon expansion through the relation $r_c=r_h(1+\epsilon)$ in eqn.(\ref{lapseRNads2}) and eq. (\ref{linu}) which leads us to the following expression
\begin{equation}\label{lrh}
l r_h=-\frac{1}{\sqrt{3}}\log[3 \epsilon]+c_1+\delta^2 c_2+\mathcal{O}(\epsilon),
\end{equation}
where, $c_1$ and $c_2$ are constants (these are listed in the Appendix (\ref{constants}) in eqs. (\ref{c1}) and (\ref{c2}) respectively). As earlier utilizing the expression for the lapse function (\ref{lapseRNads2}) and eqs. (\ref{Ainu}), (\ref{lrh}), we may evaluate the finite part of the area of the co-dimension two bulk static minimal (RT) surface as
\begin{equation}\label{shight}
{\cal A}_A^{finite} = Ll r_h^2  + {L} r_h(k_1+\delta^2 k_2)+{L}r_h\epsilon\bigg[k_3+\delta^2 (k_4+k_5 \log \epsilon)\bigg]+O[\epsilon^2],  
\end{equation}
where, $k_1$, $k_2$, $k_3$, $k_4$ and $k_5$ are constants listed in Appendix (\ref{constants}) and
from eq. (\ref{lrh}), $\epsilon$ may be expressed as
\begin{equation}\label{epsilon}
\epsilon = \frac{1}{3}\exp\left(-\sqrt{3}(l r_h - c_1 -c_2 \delta^2)\right).
\end{equation}
Hence, following our conjecture eq. (\ref {hen2}), the holographic entanglement negativity for the mixed state of the disjoint subsystems in question maybe computed as,
\begin{eqnarray}\label{ENforSQhightemp}
\mathcal{E}_\mathrm{disjoint} &=&\frac{3 }{16G_{N}^{3+1}}\bigg[L r_h\Big\{k_3(\epsilon_{1s}+\epsilon_{s2}-\epsilon_{1s2}-\epsilon_s)+\delta^2 k_4(\epsilon_{1s}+\epsilon_{s2}-\epsilon_{1s2}-\epsilon_s)\nonumber\\
&&+\delta^2 k_5\big(\epsilon_{1s}\log \epsilon_{1s}+\epsilon_{s2}\log \epsilon_{s2}-\epsilon_{1s2}\log \epsilon_{1s2}-\epsilon_s \log \epsilon_s\big)\Big\}\bigg]+\ldots,
\end{eqnarray}
where, $\epsilon_{1s}$, $\epsilon_{s2}$, $\epsilon_{1s2}$ and $\epsilon_s$ refer to the subsystems $A_1\cup A_s$, $A_s\cup A_2$, $A_1\cup A_s\cup A_2$ and $A_s$ respectively. Note that unlike the entanglement entropy the holographic entanglement negativity obtained in (\ref{ENforSQhightemp}) depends only on the area of the entangling surface in the $AdS_4/CFT_3$ scenario as all the volume dependent thermal terms cancel out. This appears to be an universal feature  of the holographic entanglement negativity \cite{Chaturvedi:2016kbk, Basak:2020bot, Chaturvedi:2016rcn, Malvimat:2018txq, Malvimat:2018ood, Jain:2017aqk,Jain:2017uhe,Jain:2017xsu,Jain:2018bai} and conforms to the quantum information theory expectations. In \cite{Jain:2017aqk,Chaturvedi:2016rcn} a similar cancellation is encountered for the $AdS_3/CFT_2$ scenario also indicating the universality of this feature.

As earlier, for a consistency check we consider the above expression for the holographic entanglement negativity in the limit of adjacent subsystems with $l_s\rightarrow a~(a\ll 1)$ and setting $l_1+l_s \to l_1$, $l_s+l_2\rightarrow l_2$ and $l_1+l_s+l_2\rightarrow l_1+l_2$  to obtain the following
\begin{eqnarray}\label{ENforSQhightempAdj}
\mathcal{E}_\mathrm{adjacent} &=&\frac{3 }{16G_{N}^{3+1}}\bigg[\Big(\frac{2L}{a}\Big) +L r_h\Big\{(k_1+\delta^2 k_2)\nonumber\\
&&+k_3(\epsilon_{1}+\epsilon_{2}-\epsilon_{12})+\delta^2 k_4(\epsilon_{1}+\epsilon_{2}-\epsilon_{12})\nonumber\\
&&+\delta^2 k_5\big(\epsilon_{1}\log \epsilon_{1}+\epsilon_{2}\log \epsilon_{2}-\epsilon_{12}\log \epsilon_{12}\big)\Big\}\bigg]+\ldots,
\end{eqnarray}
where, $\epsilon_{1}$, $\epsilon_{2}$ and $\epsilon_{12}$ refer respectively to the subsystems $A_1$, $A_2$ and $A_1\cup A_2$. The first diverging term in the above expression is obtained by utilizing eqs. \eqref{Area 3.2.1}, \eqref{A_div} and \eqref{shight} in $l_s \to a$ limit. This result matches exactly with the holographic entanglement negativity for the mixed state of adjacent subsystems with rectangular strip geometries in $CFT_3$s with a conserved charge dual to bulk non-extremal RN-$AdS_4$ black hole described in \cite{Jain:2018bai}.

\subsubsection{Large charge - high temperature}\label{section3.2.3}

Finally for the bulk non extremal black holes we consider the large charge and the high temperature regime in which
the turning point of the RT surface approaches the bulk non extremal RN-$AdS_4$ black hole horizon  with $r_c\sim r_h$ and $u_0=\frac{r_c}{r_h}\sim 1$ and we may once again adopt the near horizon approximation $r_c=r_h(1+\epsilon)$. In this case the lapse function $f(u)$ may be Taylor expanded around $u=u_0$ \cite{Chaturvedi:2016kbk} as
\begin{equation}\label{lapseRNads4}
f(u)\approx\Big(3-\frac{Q^2}{r_h^4}\Big)\Big(1-\frac{r_h}{r_c}u\Big).
\end{equation}
We denote the prefactor in the above equation as $\delta=(3-\frac{Q^2}{r_h^4})$ from now on and
employing eq. (\ref{lapseRNads4}) and (\ref{linu}) we obtain the following expressions
\begin{eqnarray}\label{lrh2}
\sqrt{\delta}lr_h=-\log[\epsilon]+k+\mathcal{O}(\epsilon), \\
\epsilon\approx \varepsilon_{ent} e^{-\sqrt{\delta}l r_h}.
\end{eqnarray}
The constants in the above expressions may be evaluated as follows
\begin{eqnarray}
\varepsilon_{ent}=e^k, \; \; \; \; k=\frac{\sqrt{\pi} \Gamma(\frac{3}{4})}{2\Gamma(\frac{5}{4})}+\sum_{n=1}^{\infty}\Big(\frac{\Gamma(n+\frac{1}{2})\Gamma(\frac{n+3}{4})}{2\Gamma(n+1)\Gamma(\frac{n+5}{4})}-\frac{1}{n}\Big).
\end{eqnarray}
We now evaluate the finite part of the area of the RT surface using eqs. (\ref{Ainu}), (\ref{lapseRNads4}) and (\ref{lrh2}) as
\begin{equation}\label{EEforLQLTemp}
{\cal A}_A^{finite}= Llr_h^2+\frac{Lr_h}{2\sqrt{\delta}}\bigg[K_1'+K_2'\epsilon+\mathcal{O}(\epsilon^2)\bigg].
\end{equation}
where the constants $K'_1$ and $K'_2$  are listed in the Appendix (\ref{appen2}).

Having obtained the area of the RT surface, we may now compute the holographic entanglement negativity for the mixed state configuration of disjoint subsystems in question through our conjecture eq. (\ref{hen2}) as
\begin{equation}\label{ENforLQhightemp}
\mathcal{E}_\mathrm{disjoint} =\frac{3}{8G_{N}^{3+1}}
\Bigg[\frac{L r_h}{\sqrt{\delta}}\Big\{K_2'(\epsilon_{1s}+\epsilon_{s2}-\epsilon_{1s2}-\epsilon_s)\Big\} \Bigg]+\ldots,
\end{equation}
where, $\epsilon_{1s}$, $\epsilon_{s2}$, $\epsilon_{1s2}$ and $\epsilon_s$ refer to the subsystems $A_1\cup A_s$, $A_s\cup A_2$, $A_1\cup A_s\cup A_2$ and $A_s$ respectively. Interestingly, as earlier
the volume dependent thermal terms entirely cancel and renders the holographic entanglement negativity to be purely dependent on the area of the entangling surface. 

Once again as a consistency check we consider the above expression in the limit of adjacent subsystems with $l_s\rightarrow a~(a\ll 1)$ and setting $l_1+l_s\rightarrow l_1$, $l_s+l_2\rightarrow l_2$ and $l_1+l_s+l_2\rightarrow l_1+l_2$ we obtain,
\begin{equation}\label{ENforLQhightempadj}
\mathcal{E}_\mathrm{adjacent} =\frac{3}{8G_{N}^{3+1}}
\Bigg[\Big(\frac{L}{a}\Big)+\frac{L r_h}{\sqrt{\delta}}\Big\{K_1'+K_2'(\epsilon_{1}+\epsilon_{2}-\epsilon_{12})\Big\} \Bigg]+\ldots,
\end{equation}
where, $\epsilon_{1}$, $\epsilon_{2}$ and $\epsilon_{12}$ once again refer to the subsystems $A_1$, $A_2$ and $A_1\cup A_2$ respectively. We have used eqs. \eqref{Area 3.2.1}, \eqref{A_div} and \eqref{EEforLQLTemp} in the adjacent limit to obtain the first diverging term in the above expression. As earlier in this regime our result exactly reproduces the holographic entanglement negativity for the mixed state of adjacent subsystems with rectangular strip geometries in $CFT_3$s with a conserved charge dual to bulk non-extremal RN-$AdS_4$ black hole obtained in \cite{Jain:2018bai}.

\subsection{Extremal RN-$\mathrm{AdS_4}$ black holes}

Having completed the case for the bulk non extremal RN-$AdS_4$ black hole geometry in this subsection we turn our attention to the holographic entanglement negativity for the bipartite zero temperature mixed state of disjoint subsystems with long rectangular strip geometries in proximity, in $CFT_3$s with a conserved charge dual to bulk extremal RN-$AdS_4$ black holes. In this context as earlier, we evaluate the integral for the area of the co dimension two bulk static minimal surface homologous to a subsystem with such long rectangular strip geometry in the dual $CFT_3$ perturbatively for the corresponding small and large charge regimes. Subsequently we utilize the results of this analysis in our holographic construction to obtain the corresponding entanglement negativity of the mixed state in question. As earlier for a consistency check we consider our results in the limit of the subsystems being adjacent to each other and demonstrate that this matches
exactly with the holographic entanglement negativity for adjacent subsystems in this case as obtained in \cite {Jain:2018bai}

\subsubsection{Small charge - extremal}

As earlier, for the extremal black holes the zero temperature $T$ leads to the following extremality condition from eq. (\ref{RNtemp}) as described in \cite{Chaturvedi:2016kbk},
\begin{equation}
r_h=\frac{\sqrt{Q}}{3^{\frac{1}{4}}}
\end{equation}
Note that in the small charge limit the turning point of the bulk RT surfaces are located far away from the black hole horizon \cite{Kundu:2016dyk} and hence we have $r_h\ll r_c$.
We may then Taylor expand the quantity $f(u)^{-\frac{1}{2}}$, where $f(u)$ is the lapse function, around $\frac{r_h}{r_c}=0$  to the leading order in $\mathcal{O}[(\frac{r_h}{r_c}u)^3]$ as \cite{Chaturvedi:2016kbk}
\begin{equation}\label{fapproxExtrm} 
f(u)^{-\frac{1}{2}}\approx1+2{\Big(\frac{r_h}{r_c} \Big)}^3u^3.
\end{equation}
Using eqs. (\ref{fapproxExtrm}), \eqref{linu} and \eqref{Ainu}, the finite part of the area of the co-dimension two bulk static minimal (RT) surface homologous to the subsystem $A$, may be computed as
\begin{equation}\label{EEareaExtrm1} 
{\cal A}_{ A}^{finite}= h_1\frac{L}{l}+h_2 r_h^3Ll^2+\mathcal{O}(r_h^4l^3),
\end{equation}
where the constants $h_1$ and $h_2$  in the above eq. (\ref{EEareaExtrm1}) have been described in
the eqs. (\ref{h1}) and (\ref{h2}). Utilizing the above result for the area of the RT surface in our construction
described in eq. (\ref{hen2}), we obtain the holographic entanglement negativity of the mixed state of disjoint subsystems
in proximity as follows
\begin{equation}\label{ENforRNSCext}
\mathcal{E}_\mathrm{disjoint} = \frac{3}{16 G_{N}^{3+1}}\Big[h_1(\frac{L}{l_1+l_s}+\frac{L}{l_s+l_2}-\frac{L}{l_1+l_s+l_2}-\frac{L}{l_s})-2h_2r_h^{3} L l_1 l_2\Big]+\ldots.
\end{equation}
As described earlier in section (\ref{small_Q_low_T}) we observe that in this case also the first term in the above equation for the holographic entanglement negativity arises from the zero temperature vacuum configuration of a $CFT_3$ with a zero charge dual to the bulk pure $AdS_4$ space time. In this case however the second term 
corresponds to the contribution due to the non zero conserved charge only as the temperature remains zero for the bulk extremal black hole.

As earlier, for a consistency check, we consider the above expression in the limit of adjacent subsystems with $l_s\rightarrow a~(a\ll 1)$ and setting $l_1+l_s\rightarrow l_1$, $l_s+l_2\rightarrow l_2$ and $l_1+l_s+l_2\rightarrow l_1+l_2$ we obtain the following expression for the holographic entanglement negativity as 
\begin{equation}
\mathcal{E}_\mathrm{adjacent} =\frac{3}{16G_{N}^{3+1}}\bigg[\Big(\frac{2L}{a}\Big)+h_1(\frac{L}{l_1}+\frac{L}{l_2}-\frac{L}{l_1+l_2})-2h_2r_h^{3} L l_1 l_2\Big]+\ldots ,
\end{equation}
where eqs. \eqref{Area 3.2.1}, \eqref{A_div} and \eqref{EEareaExtrm1} have been used. Once again our results match exactly with the holographic entanglement negativity for the mixed state of adjacent subsystems with rectangular strip geometries in $CFT_3$s with a conserved charge, dual to bulk extremal RN-$AdS_4$ black holes as described in \cite{Jain:2018bai}.

\subsubsection{Large charge - extremal}
In this limit, the turning point $r_{c}$ of the RT surface approaches the horizon radius $r_h$, so that we have $u_0=\frac{r_c}{r_h}\sim 1$. Thus we can Taylor expand the lapse function $f(u)$ around $u=u_0$ as \cite{Chaturvedi:2016kbk}
\begin{equation}\label{lapseRNads3}
f(u)\approx 6\Big(1-\frac{r_{h}}{r_{c}}u\Big)^2. 
\end{equation}
The condition $r_c\sim r_h$ allows a near horizon approximation through $r_c=r_h(1+\epsilon)$ in eq. \eqref{linu} and this leads to the following expression
\begin{equation}\label{lrhd}
l r_h=k_l+\frac{\pi}{\sqrt{6\epsilon}}+\mathcal{O}(\epsilon^{\frac{1}{2}}),
\end{equation}
where, $k_l$ is a constant which is given as
\begin{equation}\label{kl}
k_l=\sqrt{\frac{\pi}{6}}\Big[\frac{\Gamma(\frac{3}{4})}{2\Gamma(\frac{5}{4})}+\sum_{n=1}^{\infty}\Big(\frac{\Gamma(\frac{n+3}{4})}{2\Gamma(\frac{n+5}{4})}-\frac{1}{\sqrt{n}}\Big)+\zeta(\frac{1}{2})\Big]
\end{equation}
Now using eqs. \eqref{lapseRNads3}, \eqref{Ainu} and (\ref{lrhd}), we may evaluate the finite part $\mathcal{A}_A^{finite}$ of the area of the RT surface  as \cite{Chaturvedi:2016kbk}
\begin{equation}\label{EEforLQExtr}
{\cal A}_A^{finite}=Ll r_h^2 + {Lr_h}  \Big(K_1+K_2\sqrt{\epsilon}+K_3\epsilon+\mathcal{O}(\epsilon^\frac{3}{2})\Big).
\end{equation}
The constants $K_1$, $K_2$ and $K_3$ in the above eq. (\ref{EEforLQExtr}) are again listed in Appendix (\ref{appen3}).

Having obtained the expression for the area of the RT surface  we may now proceed to compute the holographic entanglement negativity for the mixed state of disjoint subsystems in question from our construction described in 
(\ref{hen2}) as follows
\begin{equation}\label{ENforLQzerotemp}
\mathcal{E}_\mathrm{disjoint} =\frac{3}{16G_{N}^{3+1}}
\bigg[L r_h\Big\{K_2(\sqrt{\epsilon_{1s}}+\sqrt{\epsilon_{s2}}-\sqrt{\epsilon_{1s2}}-\sqrt{\epsilon_{s}}) + K_3(\epsilon_{1s}+\epsilon_{s2}-\epsilon_{1s2}-\epsilon_{s})\Big\} \bigg]+\ldots,
\end{equation}
where, $\epsilon_{1s}$, $\epsilon_{s2}$, $\epsilon_{1s2}$ and $\epsilon_s$ refer to the subsystems $A_1\cup A_s$, $A_s\cup A_2$, $A_1\cup A_s\cup A_2$ and $A_s$ respectively. Note that the holographic entanglement negativity for the mixed state configuration of disjoint subsystems in question for this regime depends only on the area of the entangling surface as observed also in section (\ref{section3.2.3}) for the non-extremal case. Once again this
conforms to quantum information expectation for the entanglement negativity and is seemingly an universal feature for the corresponding dual $CFT$s.

Once again as a consistency check, we consider the above expression in the limit of adjacent subsystems with $l_s\rightarrow a~(a\ll 1)$ and setting $l_1+l_s\rightarrow l_1$, $l_s+l_2\rightarrow l_2$ and $l_1+l_s+l_2\rightarrow l_1+l_2$ to obtain,
\begin{equation}\label{ENforLQzerotempa}
\mathcal{E}_\mathrm{adjacent} =\frac{3}{16G_{N}^{3+1}}
\bigg[\Big(\frac{2L}{a}\Big)+L r_h\Big\{K_1+K_2(\sqrt{\epsilon_{1}}+\sqrt{\epsilon_{2}}-\sqrt{\epsilon_{12}}) + K_3(\epsilon_{1}+\epsilon_{2}-\epsilon_{12})\Big\} \bigg]+\ldots,
\end{equation}
where, $\epsilon_{1}$, $\epsilon_{2}$ and $\epsilon_{12}$ refer to the subsystems $A_1$, $A_2$ and $A_1\cup A_2$ respectively. Again the first diverging term in the above expression is obtained by using eqs. \eqref{Area 3.2.1}, \eqref{A_div} and \eqref{EEforLQExtr} in $l_s \to a$ limit. As earlier we observe that our result in eq. (\ref{ENforLQzerotemp}) exactly reproduces the holographic entanglement negativity for the mixed state configuration of adjacent subsystems in the $CFT_3$s with a conserved charge dual to bulk extremal RN-$AdS_4$ black hole as obtained in \cite{Jain:2018bai}.  This completes our analysis for the 
holographic entanglement negativity of the mixed state of disjoint subsystems in the $AdS_4/CFT_3$ scenario for different regimes of the charge and the temperature of the dual extremal and non extremal RN-$AdS_4$ black hole. In the following section we will extend this analysis to the generic case of $AdS_{d+1}/CFT_d$ scenario.


\section{Holographic entanglement negativity for $\mathrm{CFT_d}$ with a conserved charge dual to RN-$\mathrm{AdS_{d+1}}$ black hole}\label{sec4}

In this section we will turn our attention to the extension of  the holographic entanglement negativity construction for the $AdS_4/CFT_3$ scenario described above to a generic $AdS_{d+1}/CFT_d$ framework. In this context utilizing our construction described in section \ref{sec2}, we obtain the holographic entanglement negativity for bipartite mixed states of disjoint subsystems with long rectangular strip geometries in the $CFT_d$s with a conserved charge dual to the bulk extremal and non-extremal RN-$AdS_{d+1}$ black holes. To this end it is required to first obtain an expression for the area of the bulk RT surface corresponding to a subsystem with the geometry as described above in the dual $CFT_d$ with a conserved charge. As earlier, the area integral must be evaluated perturbatively for a generic bulk $AdS_{d+1}$ geometry. In the higher dimensional scenario it is more convenient to use the temperature $T$ and the chemical potential $\mu$ as the expansion parameters for the perturbative analysis.

\subsection{Area of the RT surface in RN-$\mathrm{AdS_{d+1}}$ black hole geometries}\label{sec4.1}

The RN-$AdS_{d+1}$ black hole metric (with the $AdS$ radius $R=1$) may be written as follows
\begin{equation}\label{henvacuum}
\begin{aligned}
 ds^2 =& \,\frac{1}{z^2} \left(- f(z) dt^2 + \frac{dz^2}{f(z)} + d\vec{x}^2 \right),\\
 f(z) = &\, 1- M z^d + \frac{(d-2)Q^2}{(d-1) } z^{2(d-1)},\\
 A_t =& \, Q \, (z_H^{d-2} - z^{d-2}),
\end{aligned}
\end{equation}
where $Q$ and $M$ are the charge and the mass of the black hole respectively. The smallest real root of the lapse function $f(z)$ gives the location of the horizon $z_H$. The chemical potential $\mu$ conjugate to the charge $Q$ is defined as
\begin{equation}\label{muforddim}
\mu \equiv \lim_{z\to 0} A_t(z) = {Q} z_H^{d-2}, 
\end{equation}
and the Hawking temperature is given by
\begin{equation}\label{Hawktempforddim}
T=-\frac{1}{4\pi}\frac{d}{dz}f(z)\bigg|_{z_H}=\frac{d}{4 \pi z_H}\left(1-\frac{(d-2)^2 Q^2 z_H^{2(d-1)}}{d(d-1)}\right). 
\end{equation}

Unlike the earlier analysis in this case it is convenient to introduce two new parameters, the effective temperature $T_{\mathrm{eff}}$ and an energy dependent parameter $\varepsilon$, which are defined in terms of the temperature $T$ and the chemical potential $\mu$ of the black hole. Here, $\varepsilon$ is a function of the expectation value of the $T_{00}$ component of the energy-momentum tensor \cite{Kundu:2016dyk} and hence describes the total energy of the dual $CFT_d$ with a conserved charge. The lapse function $f(z)$, the chemical potential $\mu$, and the temperature $T$ may be expressed in terms of  the quantity $\varepsilon$ as follows
\begin{eqnarray}
f(z)&=& 1- \varepsilon \left(\frac{z}{z_H}\right)^d + 
\left(\varepsilon-1\right) \left(\frac{z}{z_H}\right)^{2(d-1)}, \label{RNlapseddim2} \\ 
\mu&=&\frac{1}{z_H}\sqrt{\frac{ (d-1)}{(d-2)}(\varepsilon-1)}, \label{muforddim2} \\
T&=&\frac{ 2(d-1)-(d-2)\varepsilon}{4\pi z_H}. \label{Tforddim2}
\end{eqnarray}
Here $\varepsilon$ lies within the limit $1 \leq \varepsilon \leq \frac{2(d-1)}{d-2}$, and is given as
\begin{equation}\label{epsilonforddim}
\varepsilon(T,\mu)=b_0-\frac{2b_1}{1+\sqrt{1+\frac{d^2}{2\pi^2 b_0b_1}\left(\frac{\mu^2}{T^2}\right)}},
\end{equation}
with the constants $b_0$ and $b_1$ given by
\begin{equation}\label{abforddim}
b_0=\frac{2(d-1)}{d-2}\ , \qquad b_1=\frac{d}{d-2}.
\end{equation}
The number of microstates at a given temperature $T$ and chemical potential $\mu$ can then be described using the effective temperature $T_\mathrm{eff}$, defined as \cite{Kundu:2016dyk}
\begin{equation}\label{Teffforddim}
T_{\mathrm{eff}}(T,\mu) \equiv \frac{d}{4\pi z_H}=
\frac{T}{2}\left[1+\sqrt{1+\frac{d^2}{2\pi^2 b_0 b_1}\left(\frac{\mu^2}{T^2}\right)}\right].
\end{equation}

After establishing the expansion parameters for the perturbative analysis we now
proceed to the computation of the area of the RT surface corresponding to a subsystem $A$ with long rectangular strip geometry in the dual $CFT_d$ with following coordinates
\begin{equation}\label{stripforddim}
x\equiv x^1 \in \left[-\frac{l}{2},
\frac{l}{2}\right],~  x^i\in \left[-\frac{L}{2},\frac{L}{2}\right],\qquad i=2,...,d-1
\end{equation}
with $L \rightarrow \infty$. The area $\mathcal{A}_A$ of the above RT surface for the subsystem $A$ may be expressed as
\begin{equation}\label{areaforddim}
 \mathcal{A}_A=2L^{d-2}z_*^{d-1}\int_{0}^{l/2}\frac{dx}{z(x)^{2(d-1)}}=
2L^{d-2}z_*^{d-1}\int_{a}^{z_*}\frac{dz}{z^{d-1}\sqrt{f(z)[z_*^{2(d-1)}-z^{2(d-1)}]}},
\end{equation}
where $z_*$ is the turning point of the RT surface in the bulk and $a$ is a UV cut off in the dual $CFT_d$. The turning point $z_*$ is related to the length $l$ of the strip in the $x-$direction in (\ref{stripforddim}) as
\begin{equation}\label{z*forddim}
\frac{l}{2}=\int_0^{z_*}\frac{dz}{\sqrt{f(z)[(z_*/z)^{2(d-1)}-1]}}.
\end{equation}
The above integral may be expressed as a double sum as described in \cite{Kundu:2016dyk}
\begin{align}\label{z*ddim_doublesum}
l=\frac{z_*}{d-1}\sum_{n=0}^\infty \sum_{k=0}^n \frac{\Gamma\left[\frac{1}{2}+n\right]\Gamma \left[\frac{d (n+k+1)-2k}{2 (d-1)}\right]\varepsilon^{n-k} (1-\varepsilon)^k}{
\Gamma[1+n-k]\Gamma[k+1]\Gamma \left[\frac{d (n+k+2)-2k-1}{2 (d-1)}\right]}  \left(\frac{z_*}{z_H}\right)^{n d +k(d-2)} . \end{align}
The area of the RT surface in eq. (\ref{areaforddim}) may also be expressed as a double sum as \cite{Kundu:2016dyk}
\begin{align}\label{aren}
\mathcal{A}_A=& \frac{2}{d-2}\left(\frac{L}{a}\right)^{d-2}+2 \frac{L^{d-2}}{z_*^{d-2}} \left[\frac{\sqrt{\pi } \Gamma \left(-\frac{d-2}{2
(d-1)}\right)}{2 (d-1) \Gamma \left(\frac{1}{2 (d-1)}\right)}\right]\\
&+  \frac{L^{d-2}}{(d-1)z_*^{d-2}} \left[\sum_{n=1}^\infty \sum_{k=0}^n\frac{\Gamma\left[\frac{1}{2}+n\right]\Gamma \left[\frac{d (n+k-1)-2k+2}{2
(d-1)}\right]\varepsilon^{n-k} (1-\varepsilon)^k}{ \Gamma[1+n-k]\Gamma[k+1]\Gamma \left[\frac{d (n+k)-2k+1}{2 (d-1)}\right]}   \left(\frac{z_*}{z_H}\right)^{n d
+k(d-2)}\right].\nonumber
\end{align}

The exact computation of this area is, however, only possible for the zero temperature and zero chemical potential case. So we perturbatively expand the area and the turning point of the RT surface in terms of the parameters $\varepsilon (T,\mu)$ and $T_{\mathrm{eff}}(T,\mu)$, for various limits of the temperature $T$ and the chemical potential $\mu$. We then utilize the area thus obtained to compute the holographic entanglement negativity for the mixed state configuration in question.

In the next fewsubsections, we follow the procedure detailed above to compute the holographic entanglement negativity for disjoint subsystems in proximity described by long rectangular strip geometry in $CFT_d$s dual to the bulk extremal and non-extremal RN-$AdS_{d+1}$ black hole geometries. The corresponding disjoint subsystems denoted by $A_1$ and $A_2$, and the region sandwiched between them denoted by $A_s$ as depicted in fig. (\ref{fig1}), are specified by the following coordinates
\begin{eqnarray}\label{asrecd}
&x^1\in[-\frac{l_1}{2},\frac{l_1}{2}],~~~~~x^i\in[-\frac{L}{2},\frac{L}{2}],\label{a1recd}\nonumber\\
&x^1\in[-\frac{l_2}{2},\frac{l_2}{2}],~~~~~x^i\in[-\frac{L}{2},\frac{L}{2}],\label{a2recd}\\
&x^1\in[-\frac{l_s}{2},\frac{l_s}{2}],~~~~~x^i\in[-\frac{L}{2},\frac{L}{2}],\nonumber
\end{eqnarray}
respectively with $i=2,...,d-1$ and $L \to \infty$. Here $L$ denotes the length of the strip in the remaining $(d-2)$ spatial directions. The areas of the bulk RT surfaces corresponding
to the subsystems $A_1$, $A_2$ and $A_s$ may be obtained using eq. (\ref{aren}).



\subsection{Non-extremal RN-$\mathrm{AdS_{d+1}}$ black holes}
We start with the bipartite mixed state configuration of disjoint subsystems in proximity described by long rectangular strip geometries at a finite temperature $T$ in holographic $CFT_d$s with a conserved charge $Q$ dual to the bulk non-extremal RN-$AdS_{d+1}$ black holes. Here we compute the holographic entanglement negativity for the given mixed state configuration in various limits of the chemical potential $\mu$ conjugate to the charge $Q$ and the temperature $T$.

\subsubsection{Small chemical potential - low temperature}

The regime of small chemical potential and low temperature is given by $\mu l \ll 1$ and $Tl \ll 1$ where we may consider
the two limits $\mu \ll T$ and $\mu \gg T$ as follows\\

\smallskip 
\noindent{\boldmath$ (i) ~~T l\ll \mu l\ll 1$ }
\vspace{0.25cm}\\
\noindent
Here we consider the case where $T\ll\mu$ and $T l\ll \mu l \ll 1$. In this limit, we may expand the parameters $\varepsilon(T,\mu)$ and $T_{\mathrm{eff}}(T,\mu)$ given in eqs. (\ref{epsilonforddim}) and (\ref{Teffforddim}) respectively,  around $\frac{T}{\mu}=0$ to the leading order to arrive at \cite{Kundu:2016dyk}
\begin{align}
T_{\mathrm{eff}}&\approx \frac{1}{2}\left(\frac{\mu d}{\pi\sqrt{2b_0b_1}}+T\right)\label{ttllm},\\
\varepsilon &\approx b_0-\frac{2b_1\pi \sqrt{2b_0b_1}}{d}\left(\frac{T}{\mu}
\right)\label{vtllm}.
\end{align}
In this regime, the turning point of the RT surface remains far away from the horizon of the black hole \cite{Kundu:2016dyk}, i.e., $z_* \ll z_H$. The expression for $z_*$ may then be obtained from eq. (\ref{z*ddim_doublesum}) to the leading order in $(\frac{l}{z_H})^d$ as
\begin{align}
	z_*=\frac{l~\Gamma\left[\frac{1}{2(d-1)}\right]}{2 \sqrt{\pi}\Gamma\left[\frac{d}{2(d-1)}\right]} \left[1- \frac{1}{2(d+1)}\frac{2^{\frac{1}{d-1}-d} \Gamma
	\left(1+\frac{1}{2 (d-1)}\right) \Gamma \left(\frac{1}{2 (d-1)}\right)^{d+1}}{\pi^{\frac{d+1}{2}}  \Gamma
	\left(\frac{1}{2}+\frac{1}{d-1}\right)\Gamma\left(\frac{d}{2(d-1)}\right)^d}\varepsilon\left(\frac{l}{z_H}\right)^d\right.\nonumber\\
	\left.+\mathcal{O}\left(\frac{l}{z_H}\right)^{2(d-1)}\right]\ \label{turnsq}.
\end{align}
In a similar manner, the area of the RT surface in eq. (\ref{aren}) may be expanded in powers of $(\frac{l}{z_H})^d$ and re-expressed in terms of $T_{\mathrm{eff}}$ and $\varepsilon$ as \cite{Kundu:2016dyk}
\begin{equation}\label{areasq}
\mathcal{A}_A = \bigg[\frac{2}{d-2}\left(\frac{L}{a}\right)^{d-2} + 
\mathcal{S}_0 \left(\frac{L}{l}\right)^{d-2} + \varepsilon\mathcal{S}_0\mathcal{S}_1
\left( \frac{4\pi T_{\mathrm{eff}}}{d}\right)^d
 L^{d-2} l^2
\bigg]+\mathcal{O}\Big( T_{\mathrm{eff}} l\Big)^{2(d-1)},
\end{equation}
where the constants $\mathcal{S}_0$ and $\mathcal{S}_1$ which are dimension dependent are listed in Appendix \ref{B}. The holographic entanglement negativity for the mixed state configuration in question in the dual $CFT_d$ may then be obtained by utilizing our conjecture given in  eq. (\ref{hen2}) to be
\begin{align}
\mathcal{E}_\mathrm{disjoint} = \frac{3}{16G_N^{d+1}}\bigg[
\mathcal{S}_0 L^{d-2}\left(\frac{1}{(l_1+l_s)^{d-2}} +  \frac{1}{(l_2+l_s)^{d-2}} - \frac{1}{l_s^{d-2}} - \frac{1}{(l_1+l_2+l_s)^{d-2}}\right)\nonumber \\ - 
2\varepsilon\mathcal{S}_0\mathcal{S}_1
\left(\frac{4\pi T_{\mathrm{eff}}}{d} \right)^d L^{d-2} l_1 l_2\ \bigg]+\ldots.\label{EEforddimLQlowtemp}
\end{align}
The first term in above equation arises from the vacuum configuration for the mixed state of disjoint subsystems in the corresponding $CFT_d$ with zero charge and at a zero temperature, dual to the bulk pure $AdS_{d+1}$ spacetime. The second term describes the sub leading contribution due to the finite temperature and chemical potential of the black hole. 

Now as a consistency check we consider the limit of the disjoint subsystems being adjacent to each other with $l_s \rightarrow a$ in eq. (\ref{EEforddimLQlowtemp}) to arrive at
\begin{align}\label{EEforddimAdjLQlowtemp}
\mathcal{E}_\mathrm{adjacent} = \frac{3}{16G_N^{d+1}}\bigg[\frac{2}{d-2}\left(\frac{L}{a}\right)^{d-2} + 
\mathcal{S}_0 L^{d-2}\left(\frac{1}{l_1^{d-2}} +  \frac{1}{l_2^{d-2}} - \frac{1}{(l_1+l_2)^{d-2}}\right)\nonumber \\ - 
2\varepsilon\mathcal{S}_0\mathcal{S}_1
\left(\frac{4\pi T_{\mathrm{eff}}}{d} \right)^d L^{d-2} l_1l_2   \bigg]+\ldots,
\end{align}
where eq. \eqref{areasq} has been used to get the first diverging term. The above expression matches exactly with the corresponding result obtained in \cite{Jain:2018bai}. 


\smallskip 
\noindent{\boldmath$ (ii) ~~\mu l\ll T l\ll 1$ }
\vspace{0.25cm}\\
\noindent
Now we consider the scenario where $\mu \ll T$ and $\mu l\ll T l\ll 1$. In this limit $T_\mathrm{eff}(T,\mu)$ and $\varepsilon(T,\mu)$ may be Taylor expanded around $\frac{\mu}{T}=0$ as follows \cite{Kundu:2016dyk}
\begin{align}
&T_{\mathrm{eff}}(T,\mu)=T\left[1+\frac{d(d-2)^2}{16\pi^2(d-1)}
\left(\frac{\mu}{T}\right)^2+\mathcal{O}\left(\frac{\mu}{T}\right)^4\right]\ ,\label{Teffforddim2}\\
&\varepsilon(T,\mu)=1+\frac{d^2(d-2)}{16\pi^2(d-1)}\left(\frac{\mu}{T}\right)^2+
\mathcal{O}\left(\frac{\mu}{T}\right)^4\ \label{epsilonforddim2}.
\end{align}
Again the low chemical potential and the small temperature condition given by $\mu l \ll 1$ and $Tl \ll 1$ respectively implies that the turning point of the RT surface is far away from the black hole horizon \cite{Kundu:2016dyk}, $z_* \ll z_H$. So to obtain the expression for the area of the RT surface we again utilize a perturbative expansion of eq. (\ref{aren}) in  $(\frac{l}{z_H})^d$ to obtain the following \cite{Kundu:2016dyk}
\begin{equation} \label{area_mu_T_low}
~~{\cal A}_A = \Big[\frac{2}{d-2}\left(\frac{L}{a}\right)^{d-2} + 
\mathcal{S}_0 \left(\frac{L}{l}\right)^{d-2} + \varepsilon\mathcal{S}_0\mathcal{S}_1
\left( \frac{4\pi T_{\mathrm{eff}}}{d}\right)^d
 L^{d-2} l^2
\Big]+\mathcal{O}\Big( T_{\mathrm{eff}} l\Big)^{2(d-1)},
\end{equation} 
where the constants $\mathcal{S}_0$ and $\mathcal{S}_1$ are same as the ones appearing in eq. (\ref{areasq}) and are listed in Appendix \ref{B}. The expression for the area obtained in eq. (\ref{areasq}) for the case of $Tl \ll \mu l \ll 1$ is exactly same as the one obtained above for the case of $\mu l \ll T l \ll 1$. However one should be careful as the effective temperature $T_{\mathrm{eff}}$ used in the above expression for the area is given in eq. (\ref{Teffforddim2}) and is different from the one used in eq. (\ref{areasq}).

Utilizing our conjecture in eq. (\ref{hen2}), we may now obtain the holographic entanglement negativity for the corresponding mixed state configuration of disjoint subsystems in proximity with long rectangular strip geometry as
\begin{align}
\mathcal{E}_\mathrm{disjoint} = \frac{3}{16G_N^{d+1}}\Big[ \mathcal{S}_0 L^{d-2}\left(\frac{1}{(l_1+l_s)^{d-2}} +  \frac{1}{(l_s+l_2)^{d-2}} -\frac{1}{l_s^{d-2}} -  \frac{1}{(l_1+l_s+l_2)^{d-2}}\right)\nonumber 
\\ - 2\varepsilon\mathcal{S}_0\mathcal{S}_1 \left(\frac{4\pi T_{\mathrm{eff}}}{d} \right)^d L^{d-2} l_1 l_2\   \Big]+\ldots.\label{HENmllt}
\end{align}
Note that here also the first term in the above equation is what one would obtain for the case of disjoint subsystems in proximity in a $CFT_d$ with zero charge and at zero temperature, dual to the bulk pure $AdS_{d+1}$ geometry and
the second term is the contribution due to the finite temperature and the chemical potential of the black hole. The expression obtained above in eq. (\ref{HENmllt}) exactly matches in the adjacent limit ($l_s \to a$) with the corresponding results obtained in \cite{Jain:2018bai}.

\subsubsection{Small chemical potential - high temperature}
In this section, we discuss the limit of small chemical potential and high temperature, i.e., $T \gg \mu$ and $Tl \gg 1$. Here we Taylor expand the parameters $T_{\mathrm{eff}} (T,\mu)$ and $\varepsilon (T,\mu)$ around $\frac{\mu}{T}=0$, to the leading order as given by the eq. (\ref{Teffforddim2}) and (\ref{epsilonforddim2}). But in contrast to the previous case, in this limit the RT surface goes deep into the bulk approaching the black hole horizon but never penetrating it \cite{Kundu:2016dyk, Hubeny:2012ry}, i.e., $z_* \sim z_H$. Hence, for the area of the RT surface in this limit, we take $z_*=z_H(1-\epsilon)$, where $\epsilon \ll 1$, in eq. (\ref{areaforddim}) and perform the integration to the leading order in $\epsilon$ to obtain the following
\begin{align}\label{AreaforddimSQhightemp}
\mathcal{A}_A=&\Big[\frac{2}{d-2}\left(\frac{L}{a}\right)^{d-2}+L^{d-2} l \left(\frac{4 \pi T_{\mathrm{eff}}}{d}\right)^{d-1}+L^{d-2}\left(\frac{4 \pi T_{\mathrm{eff}}}{d}\right)^{d-2}\gamma_d\left(\frac{\mu}{T}\right)
\nonumber  \\ &+\epsilon L^{d-2} \left(\frac{4 \pi T_{\mathrm{eff}}}{d}\right)^{d-2}\Big\{(d-1)l \left( \frac{4 \pi T_{\mathrm{eff}}}{d}\right)+(d-2)\gamma_d\left(\frac{\mu}{T}\right)\Big\}+\mathcal{O}(\epsilon^2)\Big],
\end{align}
where the function $\gamma_d(\frac{\mu}{T})$ is listed in the Appendix \ref{B}. Also, the parameter $\epsilon$ can be expressed in terms of the length $l$ of the subsystem $A$ in the $x^1-$direction in eq. (\ref{asrecd}) as follows \cite{Kundu:2016dyk}
\begin{equation}\label{smallepsilonford}
\epsilon=C_d  \, \mathrm{exp}\Big(-\alpha_d(\varepsilon) T_\mathrm{eff} l\Big),
\end{equation}
where $C_d$ is a dimension dependent $\mathcal{O}(1)$ numerical constant and
\begin{equation}
\alpha_d(\varepsilon)=\frac{2\pi}{d}\sqrt{2(d-1)\{d(2-\varepsilon)+2\varepsilon-2\}}.
\end{equation}
The holographic entanglement negativity for the given disjoint subsystems may then be obtained using our conjecture in eq. (\ref{hen2}) to be
\begin{align}\label{ENforddim3}
\mathcal{E}_\mathrm{disjoint} = \frac{3 L^{d-2}}{16G_N^{d+1}} \left(\frac{4 \pi T_{\mathrm{eff}}}{d}\right)^{d-2}\bigg[ 4 \pi T_{\mathrm{eff}}\frac{d-1}{d}\Big\{\epsilon_{1s} (l_1+l_s)+\epsilon_{s2} (l_s+l_2)-\epsilon_s l_s 
\nonumber \\-\epsilon_{1s2} (l_1+l_s+l_2) \Big\}+(d-2)\gamma_d\left(\frac{\mu}{T}\right)\Big\{\epsilon_{1s}+\epsilon_{s2}-\epsilon_{s}-\epsilon_{1s2} \Big\} \bigg]+\ldots,
\end{align}
where $\epsilon_{1s}$, $\epsilon_{s2}$, $\epsilon_s$ and $\epsilon_{1s2}$ correspond to the subsystems $A_1\cup A_s$, $A_s\cup A_2$, $A_s$ and $A_1\cup A_s\cup A_2$, respectively. Note here that the above expression for the holographic entanglement negativity for disjoint subsystems is completely independent of the volume dependent thermal contributions, i.e., terms proportional to $L^{d-2} l$ in the expression for area given in eq. (\ref{AreaforddimSQhightemp}) cancel out completely, conforming to the standard quantum information theory expectations. This provides a strong substantiation for our conjecture.

As another consistency check, we take the adjacent limit of the subsystems with $l_s \to a$ in eq. (\ref{ENforddim3}) to obtain
\begin{align}\label{ENadjforddimLQHT}
	\mathcal{E}_\mathrm{adjacent} =& \frac{3}{16G_N^{d+1}}\bigg[\frac{2}{d-2}\left(\frac{L}{a}\right)^{d-2}+
	L^{d-2}\left( \frac{4\pi T_{\mathrm{eff}}}{d}\right)^{d-2}\gamma_{d}\left(\frac{\mu}{T}\right)  +L^{d-2}(d-1)\left( \frac{4\pi T_\mathrm{eff}}{d}\right)^{d-2}
 \nonumber	\\
	&\qquad\qquad \bigg\{\frac{4\pi T_\mathrm{eff}}{d}\Big( \epsilon_1 l_1 + \epsilon_2 l_2 - \epsilon_{12} l_{12}\Big) + \gamma_d\left(\frac{\mu}{T}\right) \frac{d-2}{d-1}\Big( \epsilon_1+ \epsilon_2 - \epsilon_{12}\Big)\bigg\}\bigg]+\ldots.
\end{align}
Here, we have used eq. \eqref{AreaforddimSQhightemp} in the adjacent limit to obtain the first diverging term. The expression obtained above matches up to $\mathcal{O}(\epsilon^0)$ exactly with the results given in \cite{Jain:2018bai} for the corresponding adjacent subsystems. The $\mathcal{O}(\epsilon)$ terms can be considered as the higher order correction to the results in \cite{Jain:2018bai}


\subsubsection{Large chemical potential - low temperature}

We now proceed to the limit of large chemical potential and low temperature given by the conditions $\mu l \gg 1$ and $\mu \gg T$. The last limit allows us to modify the parameters $T_\mathrm{eff}(T,\mu)$ and $\varepsilon(T,\mu)$ by expanding them around $\frac{T}{\mu}=0$ like earlier to obtain eq. (\ref{ttllm}) and (\ref{vtllm}) respectively. The  first limit $\mu l \gg 1$ implies that we are again working in the limit where the turning point of the RT surface is approaching the black hole horizon \cite{Kundu:2016dyk, Hubeny:2012ry}, i.e., $z_* \sim z_H$. Hence, we may substitute $z_*=z_H(1-\epsilon)$ with $\epsilon \ll 1$ in eq. (\ref{areaforddim}) and compute the area as follows
\begin{align}\label{AreaforddimLQhightemp}
\mathcal{A}_A=&\Big[\frac{2}{d-2}\left(\frac{L}{a}\right)^{d-2}+L^{d-2} l \left(\frac{4 \pi T_{\mathrm{eff}}}{d}\right)^{d-1}+L^{d-2}\left(\frac{4 \pi T_{\mathrm{eff}}}{d}\right)^{d-2}\Big\{N_0 + N_1(b_0-\varepsilon)\Big\}
 \nonumber \\ &+\epsilon L^{d-2} \left(\frac{4 \pi T_{\mathrm{eff}}}{d}\right)^{d-2}\Big\{4 \pi T_{\mathrm{eff}} l \frac{d-1}{d}+(d-2)\Big(N_0 + N_1(b_0-\varepsilon)\Big)\Big\}+\mathcal{O}(\epsilon^2)\Big],
\end{align}
where $\epsilon$ can be written in terms of the length of the subsystem considered as given in eq. (\ref{smallepsilonford}) and the numerical constants $N_0$ and $N_1$  are listed in Appendix \ref{B}. Now we may utilize our conjecture in eq. (\ref{hen2}) to compute the holographic entanglement negativity for the disjoint subsystems in question to the leading order as follows 
\begin{align}\label{EN_HTHQ_dis}
\mathcal{E}_\mathrm{disjoint} =& \frac{3 L^{d-2}}{16G_N^{d+1}}\left(\frac{4 \pi T_{\mathrm{eff}}}{d}\right)^{d-2}\bigg[4 \pi T_{\mathrm{eff}} \frac{d-1}{d}\Big\{\epsilon_{1s} (l_1+l_s)+\epsilon_{s2} (l_s+l_2)-\epsilon_s l_s 
\nonumber \\ &-\epsilon_{1s2} (l_1+l_s+l_2) \Big\}+(d-2)\Big\{N_0 + N_1(b_0-\varepsilon)\Big\}\Big(\epsilon_{1s}+\epsilon_{s2}-\epsilon_{s}-\epsilon_{1s2} \Big) \bigg]+\ldots,
\end{align}
where $\epsilon_{1s}$, $\epsilon_{s2}$, $\epsilon_s$ and $\epsilon_{1s2}$ correspond the subsystems $A_1 \cup A_s$, $A_s \cup A_2$, $A_s$ and $A_1 \cup A_s \cup A_2$, respectively. Similar to the previous subsections, notice that in the expression for the holographic entanglement negativity obtained above the volume dependent thermal terms cancel out. And in the limit of the subsystems being adjacent ($l_s \rightarrow a$), eq. (\ref{EN_HTHQ_dis}) reduces to 
\begin{align}\label{ENadjforddimHQHT}
\mathcal{E}_\mathrm{adjacent} = \frac{3}{16G_N^{d+1}}\bigg[ & \frac{2}{d-2}\left(\frac{L}{a}\right)^{d-2}
+ L^{d-2}\left( \frac{4\pi T_{\mathrm{eff}}}{d}\right)^{d-2}\Big\{N_0 + N_1(b_0-\varepsilon)\Big\} \nonumber\\
& +L^{d-2}(d-1)\left(\frac{4\pi T_\mathrm{eff}}{d}\right)^{d-2} \nonumber\bigg\{\frac{4\pi T_\mathrm{eff}}{d}\Big( \epsilon_1 l_1 + \epsilon_2 l_2 - \epsilon_{12} l_{12}\Big)\\
& + \frac{d-2}{d-1} \Big\{N_0 + N_1(b_0-\varepsilon)\Big\} \Big( \epsilon_1+ \epsilon_2 - \epsilon_{12}\Big)\bigg\}
\bigg] + \ldots,
\end{align}
where eq. \eqref{AreaforddimLQhightemp} has been used to obtain the first diverging term. Note that in the above expression for the holographic entanglement negativity the $\mathcal{O}(\epsilon^0)$ terms are identical to the corresponding adjacent subsystems result described in \cite{Jain:2018bai} and similar to the previous case, $\mathcal{O}(\epsilon)$ terms can be considered as higher order corrections to the results in \cite{Jain:2018bai}.


\subsection{Extremal RN-$\mathrm{AdS_{d+1}}$ black holes}

Having described in detail the holographic entanglement negativity for the mixed state configuration of disjoint subsystems in proximity, in $CFT_d$s with conserved charge dual to  bulk non extremal RN-$AdS_{d+1}$ black holes we now turn our attention to $CFT_d$s at zero temperature dual to bulk extremal RN-$AdS_{d+1}$ black holes. The relevant parameters for the extremal black hole in this case are given as \cite{Kundu:2016dyk}
\begin{align}
Q^2 \, =& \,  \frac{d(d-1)}{(d-2)^2z_H^{2(d-1)}},\label{Qext}\\
\varepsilon \, =& \, b_0,\\
\mu \, = & \, \frac{1}{z_H}\sqrt{\frac{b_0 \, b_1}{2}}=\frac{1}{z_H}\sqrt{\frac{d (d-1)}{(d-2)^2}},\\
T_{\mathrm{eff}} \, =& \, \frac{\mu \, d}{2\pi\sqrt{2 \, b_0 \, b_1}},
\end{align}
where Q is the charge of the extremal RN-$AdS_{d+1}$ black hole and $\mu$ is its conjugate chemical potential. Similar to the non-extremal case, in the following subsections we  perturbatively compute the area of the RT surface homologous to
a subsystem with long rectangular strip geometry in the dual $CFT_d$ for various regimes of the chemical potential $\mu$. We then utilize our construction to obtain the holographic entanglement negativity for the mixed state configuration of disjoint subsystems in proximity under consideration.

\subsubsection{Small chemical potential}
We start with the small chemical potential case given by $\mu l \ll 1$, for which the RT surface remains close to the boundary \cite{Kundu:2016dyk}. This allows us to expand the area of this RT surface given in eq. (\ref{aren}) perturbatively to the leading order in $(\frac{l}{z_H})^d$ which may be expressed in terms of the chemical potential $\mu$ as follows \cite{Kundu:2016dyk}
\begin{equation}\label{EEforddimLQExtr}
{\cal A}_A =\bigg[\frac{2}{d-2}\left(\frac{L}{a}\right)^{d-2} + 
\mathcal{S}_0 \left(\frac{L}{l}\right)^{d-2} + \mathcal{S}_0 \, \mathcal{S}_1 \frac{2 (d-1)}{d-2}
\left(\frac{(d-2)\mu}{\sqrt{d(d-1)}} \right)^d L^{d-2} l^2
+\mathcal{O}\left((\mu l)^{2(d-1)}\right)\bigg],
\end{equation} 
where the constants $\mathcal{S}_0$ and $\mathcal{S}_1$ are same as the ones encountered in eq. (\ref{areasq}) and are listed in Appendix \ref{B}. The holographic entanglement negativity for the disjoint subsystems in question may then be obtained by utilizing our conjecture in eq. (\ref{hen2}) to be as follows
\begin{align}\label{ENforddimExtLQ}
\mathcal{E}_\mathrm{disjoint} =  \frac{3}{16G_N^{d+1}}\bigg[ \mathcal{S}_0 L^{d-2}\left(\frac{1}{(l_1+l_s)^{d-2}} +  \frac{1}{(l_s+l_2)^{d-2}} -\frac{1}{l_s^{d-2}} - \frac{1}{(l_1+l_s+l_2)^{d-2}}\right)
\nonumber \\ - 2 \, \varepsilon \, \mathcal{S}_0 \, \mathcal{S}_1 \, \frac{2(d-1)}{d-2}
\left(\frac{(d-2)\mu}{\sqrt{d(d-1)}} \right)^d L^{d-2} l_1l_2\ \bigg]+\ldots.
\end{align}
Note that the first term in the above expression is the vacuum contribution for the disjoint subsystems in a $CFT_d$ with zero charge dual to the bulk pure $AdS_{d+1}$ geometry, and the last term describes the contribution due to the non zero conserved charge. As a consistency check, we consider the adjacent limit ($l_s \to a$) of our result in eq. (\ref{ENforddimExtLQ}) along with eq. \eqref{EEforddimLQExtr} to arrive at
\begin{align}\label{ENadjforddimExtLQ}
\mathcal{E}_\mathrm{adjacent} = \frac{3}{16G_N^{d+1}}\bigg[\frac{2}{d-2}\left(\frac{L}{a}\right)^{d-2} + 
\mathcal{S}_0 L^{d-2}\left(\frac{1}{l_1^{d-2}} +  \frac{1}{l_2^{d-2}} - \frac{1}{(l_1+l_2)^{d-2}}\right)\nonumber \\ - 2 \, \varepsilon \, \mathcal{S}_0 \, \mathcal{S}_1 \, \frac{2(d-1)}{d-2}
\left(\frac{(d-2)\mu}{\sqrt{d(d-1)}} \right)^d L^{d-2} l_1l_2   \bigg]+\ldots.
\end{align}
which exactly matches with the corresponding expression obtained in \cite{Jain:2018bai}

\subsubsection{Large chemical potential}
Finally we consider the regime of large chemical potential given by $\mu l \gg 1$, for which the RT surface penetrates deep inside the bulk, with the turning point $z_*$ approaching the black hole horizon $z_H$ \cite{Kundu:2016dyk, Hubeny:2012ry}. In this case we may consider $z_*=z_H \, (1-\epsilon)$ with $\epsilon \ll 1$ in eq. (\ref{areaforddim}) and expand the area integral perturbatively in terms of $\epsilon$ which may be expressed in terms of the chemical potential as follows
\begin{align}\label{AreaforddimHQzerotemp}
{\cal A}_A = \bigg[\frac{2}{d-2}\left(\frac{L}{a}\right)^{d-2} +  L^{d-2} l \mu^{d-1}\left( \frac{d-2}{\sqrt{d(d-1)}}\right)^{d-1} + L^{d-2} \, N(b_0) \left( \frac{d-2}{\sqrt{d(d-1)}} \right)^{d-2}\mu^{d-2}
\nonumber \\ +\epsilon L^{d-2} \mu^{d-2} \left(\frac{d-2}{\sqrt{d(d-1)}}\right)^{d-2} \Big\{ l \, (d-1) \, \mu \frac{d-2}{\sqrt{d(d-1)}} +(d-2) \, N(b_0)\Big\} +\mathcal{O}(\epsilon^2)\bigg], 
\end{align}
where the function $N(\varepsilon)$, given in Appendix \ref{B}, is calculated at $\varepsilon=b_0$. The parameter $\epsilon$ appearing in the above expression for the area may be described in terms of the length $l$ of the subsystem $A$ as follows \cite{Kundu:2016dyk}
\begin{equation}
\epsilon \, = \, \frac{d}{2 \pi^2 \, (d-1)^2 \, l^2 \, T_{\mathrm{eff}}^2} \, = \, \frac{8}{(d-1) \, (d-2)^2 \, \mu ^2 \, l^2}.
\end{equation}
By substituting the area of the RT surface obtained in eq. (\ref{AreaforddimHQzerotemp}) in our conjecture described in eq. (\ref{hen2}), we may obtain the holographic entanglement negativity for the  mixed state configuration in question as
\begin{align}\label{ENforddimExtHQ}
\mathcal{E}_\mathrm{disjoint} = \frac{3  L^{d-2} \mu^{d-2}}{16 G_N^{d+1}} \, \left(\frac{d-2}{\sqrt{d(d-1)}}\right)^{d-2} \bigg[  \mu (d-2)\sqrt{\frac{d-1}{d}} \Big\{\epsilon_{1s} \, (l_1+l_s) + \epsilon_{s2} \, (l_s+l_2) \nonumber \\ 
-\epsilon_{s} \, l_s - \epsilon_{1s2} \, (l_1+l_s+l_2)\Big\} +(d-2) \, N(b_0) \, \Big\{\epsilon_{1s}+\epsilon_{s2}-\epsilon_{s}-\epsilon_{1s2} \Big\} \bigg]+\ldots,
\end{align}
where, similar to previous subsections, $\epsilon_{1s}$, $\epsilon_{s2}$, $\epsilon_s$ and $\epsilon_{1s2}$ correspond to the subsystems $A_1 \cup A_s$, $A_s \cup A_2$, $A_s$ and $A_1 \cup A_s \cup A_2$, respectively. Again, one should note here that the volume dependent thermal term present in the expression of area cancel out entirely. Implementing the adjacent limit ($l_s \rightarrow a$), we arrive at the expression for the holographic entanglement negativity for the adjacent subsystems as
\begin{align}\label{ENadjforddimExtHQ}
	\mathcal{E}_{\mathrm{adjacent}} = \frac{3}{16 G_N^{d+1}}\bigg[ & \frac{2}{d-2}\left(\frac{L}{a}\right)^{d-2} +  
	L^{d-2}N(b_0)\left(\frac{\mu(d-2)}{\sqrt{d(d-1)}}\right)^{d-2} \nonumber\\
	&+L^{d-2}\left(\frac{\mu(d-2)}{\sqrt{d(d-1)}}\right)^{d-2}\Big\{ \mu (d-2)\, \sqrt{\frac{d-1}{d}}\Big( \epsilon_1 \, l_1 + \epsilon_2 \, l_2 - \epsilon_{12} \, l_{12}\Big) \nonumber\\
	&  +(d-2) \, N(b_0) \Big( \epsilon_1+ \epsilon_2 - \epsilon_{12}\Big)\Big\}
	\bigg]+\ldots,
\end{align}
where eq. \eqref{AreaforddimHQzerotemp} has been used to obtain the first diverging term. Similar to previous cases the $\mathcal{O} (\epsilon^0)$ terms in eq. (\ref{ENadjforddimExtHQ}) match exactly with the corresponding result obtained in \cite{Jain:2018bai}. Again the  $\mathcal{O} (\epsilon)$ terms are to be interpreted as higher order corrections. This completes our computation of the holographic entanglement negativity for bipartite mixed states of disjoint subsystems in proximity with long rectangular strip geometries in $CFT_d$s with a conserved charge dual to bulk RN-$AdS_{d+1}$ black holes

\section{Summary and conclusion}\label{sec5}

To summarize, we have extended a holographic construction to compute the entanglement negativity for bipartite mixed state configurations of two disjoint subsystems in proximity to $CFT_d$s at zero and finite temperature dual to bulk extremal and non extremal RN-$AdS_{d+1}$ black holes respectively. Our proposal involved an algebraic sum of the areas of RT surfaces homologous to certain appropriate combinations of the subsystems in question. Using our construction we have computed the holographic entanglement negativity for such mixed states described by two disjoint subsystems with long rectangular strip geometries in dual $CFT_d$s with a conserved charge. In this context we have first analysed the $AdS_4/CFT_3$ scenario which required a perturbative evaluation for the area of the corresponding RT surfaces for various limits of certain parameters for the bulk RN-$AdS_4$ black hole geometry. In this context we first computed the holographic entanglement negativity for finite temperature mixed states of disjoint subsystems in a $CFT_3$ dual to bulk non-extremal RN-$AdS_4$ black holes. For the regime of small charge and low temperature the holographic entanglement negativity included a leading zero temperature contribution arising from the pure $AdS_4$ background and sub leading terms involving the charge and the temperature. We then analyzed the other regimes of  small charge and high temperature and that of large charge and high temperature. Interestingly in all these regimes the holographic entanglement negativity was cut off independent. Furthermore, our results were dependent only on the area of the entangling surfaces as the volume dependent thermal terms canceled out from the final expression. This characteristic was consistent with the quantum information theory expectation for the entanglement negativity as an upper bound on the distillable entanglement which precisely excludes the contributions from the thermal correlations. As a consistency check we have considered our results in the limit of adjacent subsystems for all of the above cases and they matched exactly with the corresponding results in the literature. All of these constitute strong substantiations for our holographic construction. 

Subsequently we extended our analysis to the case of the zero temperature mixed state of disjoint subsystems in the $CFT_3$ dual to bulk extremal RN-$AdS_4$ black holes. Once again we obtained the holographic entanglement negativity for various regimes of the parameters as described above. For the regime of small charge, the holographic entanglement negativity involved a contribution from the zero charge sector of the $CFT_3$ dual to the bulk pure $AdS_4$ space time and higher order contributions dependent on the charge $Q$ of the bulk extremal RN-$AdS_4$ black hole. The large charge regime was also analyzed in a similar way to compute the corresponding holographic entanglement negativity of the mixed state under consideration. As earlier for both the cases the results in the adjacent subsystem limit matched exactly with those in the literature and the holographic entanglement negativity was cut off independent and only dependent on the area of the corresponding entangling surfaces as expected.
 
Following the above analysis for the $AdS_4/CFT_3$ scenario we extended our holographic construction to a generic
$AdS_{d+1}/CFT_d$ framework. The perturbative evaluation for the area of the RT surfaces in this case was facilitated through the introduction of two new parameters which were dependent on the chemical potential and the temperature of the
bulk RN-$AdS_{d+1}$ black holes. The results for the holographic entanglement negativity for the various regimes of these parameters followed a similar pattern as those for the $AdS_4/CFT_3$ scenario, depending only on the area of the entangling surfaces. In all the cases the results once again matched exactly with those for the case of adjacent subsystems in the literature for the appropriate limit and as earlier served as important consistency checks. Note that, in the literature the holographic entanglement negativity for two adjacent subsystems had been computed upto the  leading order for the cases, where the RT surface approached the horizon, whereas in this article we computed the same for two disjoint subsystems upto the first sub leading term which was dependent on the lengths of the corresponding subsystems. Furthermore 
as earlier these results in the limit of adjacent subsystems matched exactly with those described in the literature upto the leading order.

Our holographic construction for the entanglement negativity of mixed states described by disjoint subsystems in 
$CFT_d$s with a conserved charge dual to bulk non extremal and extremal RN-$AdS_{d+1}$ black holes provides significant substantiation for the universality of our proposal which was earlier applied to $CFT_d$s dual to bulk 
Schwarzschild-$AdS_{d+1}$ black holes. In both cases our results are in conformity with quantum information theory expectations and in the appropriate limit matches with the results for adjacent subsystems in the literature. However 
our results are valid only for subsystems with a specific long rectangular strip geometry. The application of our construction to generic subsystem geometries is a non trivial open issue which needs attention. We should mention here
that a bulk proof for our holographic construction may be inferred from certain recent developments regarding novel replica symmetry breaking saddles for the gravitational path integral for spherical entangling surfaces. However a generalization to other subsystem geometries such as the subsystems with long rectangular strip geometries used in our examples is still an open issue. Furthermore as discussed in the introduction an alternative holographic construction for the entanglement negativity of bi partite states in $CFT_d$s from the backreacted minimal entanglement wedge cross section was recently developed where the backreaction parameter was an overall dimension dependent numerical constant which could be explicitly evaluated for spherical entangling surfaces. In this light of these developments our results should also be modified by an overall dimension dependent numerical constant, however an explicit evaluation of this factor for the subsystem geometries used in this article is still a non trivial open problem. Note that such a modification by an overall numerical factor will still preserve all our results and the consequent physics. It would be extremely significant
to further investigate the extension for the bulk proof mentioned above to generic subsystem geometries and also explicitly evaluate the backreaction factor for such subsystems. These significant but non trivial investigations are expected to lead to new directions and insight into holographic entanglement for mixed states which may have important applications to issues of quantum gravity as well as strongly coupled condensed matter systems. We hope to return to these exciting issues in the near future.

\section{Acknowledgment}
We would like to thank Vinay Malvimat and Debarshi Basu for useful discussions and suggestions.


\begin{appendices}

	\section{Non-extremal  and extremal RN-$\mathrm{AdS_4}$}\label{appenA}
	
	\subsection{Non-extremal RN-$\mathrm{AdS_4}$ (Small charge - high temperature)}\label{appen1}
	
	The constants  $k_1$, $k_2$, $k_3$, $k_4$ and $k_5$  in the eq. (\ref{shight}) are given as follows
	\begin{eqnarray}\label{constants}
	k_1 &=& \sum_{n=1}^\infty\bigg(\frac{1}{3n-1} \frac{\Gamma(n+\frac{1}{2})}
	{\Gamma(n+1)}\frac{\Gamma(\frac{3n+3}{4})}{\Gamma(\frac{3n+5}{4})}-\frac{2}{3\sqrt{3}n^2}\bigg) + \frac{\pi^2}{9\sqrt{3}}
	+\frac{\sqrt{\pi} \Gamma(-\frac{1}{4})}{\Gamma(\frac{1}{4})},\label{k1}\\
	k_2 &=& \frac{3\pi}{8}-\frac{3 \Gamma(\frac{3}{2})\Gamma(\frac{7}{4})}{\Gamma(\frac{9}{4})}+3\sum_{n=1}^\infty\bigg(\frac{1}{3n+2} \frac{\Gamma(n+\frac{3}{2})}{\Gamma(n+1)}\frac{\Gamma(\frac{3n+6}{4})}{\Gamma(\frac{3n+8}{4})}-\frac{1}{3\sqrt{3}n}\bigg)\nonumber\\
	&-& 3\sum_{n=1}^\infty\bigg(\frac{2}{3n+3} \frac{\Gamma(n+\frac{3}{2})}{\Gamma(n+1)}\frac{\Gamma(\frac{3n+7}{4})}{\Gamma(\frac{3n+9}{4})}-\frac{2}{3\sqrt{3}n}\bigg)\label{k2},\\
	k_3&=&\frac{-2}{\sqrt{3}} + \frac{\pi^2}{9\sqrt{3}},\label{k3}\\
	k_4 &=& \frac{2}{\sqrt{3}}-\frac{2}{\sqrt{3}}\log[3]+\frac{3\sqrt{\pi}\Gamma(\frac{3}{2})\Gamma(\frac{7}{4})}{\Gamma(\frac{9}{4})},\label{k4}\\ 
	k_5 &=& \frac{-2}{\sqrt{3}}\label{k5}.
	\end{eqnarray}
	The constants $c_1$ and $c_2$ appearing  in  the eq. (\ref{epsilon})  are given as follows
	\begin{eqnarray}
	c_1 &=& \frac{\sqrt{\pi}}{2}\frac{\Gamma(\frac{3}{4})}{\Gamma(\frac{5}{4})}+ \sum_{n=1}^\infty\bigg( \frac{\Gamma(n+\frac{1}{2})}{2\Gamma(n+1)}\frac{\Gamma(\frac{3n+3}{4})}{\Gamma(\frac{3n+5}{4})}-\frac{1}{\sqrt{3}n}\bigg),\label{c1}\\
	c_2 &=& \frac{1}{\sqrt{3}}-\frac{3}{2}\sum_{n=0}^\infty \bigg( \frac{\Gamma(n+\frac{3}{2})}{\Gamma(n+1)}\frac{\Gamma(\frac{3n+6}{4})}{\Gamma(\frac{3n+8}{4})}-\frac{2}{\sqrt{3}}\bigg) +\frac{3}{2}\sum_{n=0}^\infty\bigg( \frac{\Gamma(n+\frac{3}{2})}{\Gamma(n+1)}\frac{\Gamma(\frac{3n+7}{4})}{\Gamma(\frac{3n+9}{4})}-\frac{2}{\sqrt{3}}\bigg).\label{c2}
	\end{eqnarray}
	
	\subsection{Non-extremal RN-$\mathrm{AdS_4}$ (Large charge - high temperature)}\label{appen2}
	
	The constants $K'_1$ and $K'_2$  in the eq. (\ref{EEforLQLTemp}) are given as follows
	\begin{eqnarray}
	K_1'&=&-\frac{2\sqrt{\pi}\Gamma(\frac{3}{4})}{\Gamma(\frac{1}{4} )}+\frac{\log[4]-10}{8} +\frac{1}{2}\sum_{n=2}^\infty\bigg(\frac{1}{n-1}\frac{\Gamma(n+\frac{1}{2})}{\Gamma(n+1)}\frac{\Gamma(\frac{n+3}{4})}{\Gamma(\frac{n+5}{4})}-\frac{2}{n^2}\bigg)+\frac{\pi^2}{6},\label{K'1}\\
	K_2'&=& \frac{\pi^2 }{6}- \frac{3}{2}\label{K'2}.
	\end{eqnarray}

	\subsection{Extremal RN-$\mathrm{AdS_4}$ (Large charge)}\label{appen3}
	
	The constants $K_1$, $K_2$ and $K_3$ in the eq. (\ref{EEforLQExtr}) are given as follows
	\begin{eqnarray}
	K_1 &=& \frac{2}{\sqrt{6}}\bigg[-2\frac{\sqrt{\pi } \Gamma (\frac{3}{4})}{ \Gamma (\frac{1}{4})}+\frac{\log[4]}{4}-\frac{1+2\sqrt{\pi}}{2}+ \sqrt{\pi } \zeta \left(\frac{3}{2}\right)\nonumber\\
	&+&\frac{\sqrt{\pi }}{2}\sum^\infty_{n=2}
	(\frac{1}{n-1}\frac{\Gamma (\frac{n+3}{4})}{ \Gamma (\frac{n+5}{4})}-\frac{2}{n\sqrt{n}})\bigg],\label{K1} \\
	K_2 &=& -\frac{2\pi}{\sqrt{6}},\label{K2} \\
	K_3&=&\frac{2}{\sqrt{6}}\bigg[1-\sqrt{\pi}+
	\sqrt{\pi }  \zeta \left(\frac{3}{2}\right)\bigg].\label{K3}
	\end{eqnarray}
	
	
	\section{Non-extremal  and extremal RN-$\mathrm{AdS_{d+1}}$ black holes}\label{B}
	
	\subsection{Non-extremal RN-$\mathrm{AdS_{d+1}}$ black holes (Small chemical potential - low temperature)}\label{B1}
	
	The constants $\mathcal{S}_0$ and $\mathcal{S}_1$ in eq. (\ref{areasq}) which depend solely on the spacetime dimension are given as follows
	\begin{align}
	\mathcal{S}_0=& \frac{2^{d-2} \pi ^{\frac{d-1}{2}} \Gamma 
		\left(-\frac{d-2}{2 (d-1)}\right) }{(d-1) \Gamma \left(\frac{1}{2 (d-1)}\right)} \left(\frac{\Gamma
		\left(\frac{d}{2 (d-1)}\right)}{\Gamma \left(\frac{1}{2 (d-1)}\right)}\right)^{d-2}, \label{s0} \\
	\mathcal{S}_1=& \frac{\Gamma \left(\frac{1}{2 (d-1)}\right)^{d+1}2^{-d-1} 
		\pi ^{-\frac{d}{2}}}{\Gamma \left(\frac{d}{2(d-1)}\right)^d\Gamma
		\left(\frac{1}{2}+\frac{1}{d-1}\right)} \left(\frac{\Gamma \left(\frac{1}{d-1}\right) }{\Gamma \left(-\frac{d-2}{2 (d-1)}\right)}+\frac{2^{\frac{1}{d-1}} (d-2)
		\Gamma \left(1+\frac{1}{2 (d-1)}\right) }{\sqrt{\pi } (d+1)}\right).\label{s1}
	\end{align}

	\subsection{Non-extremal RN-$\mathrm{AdS_{d+1}}$ black holes (Small chemical potential - high temperature)}\label{B2}
	
	The function $\gamma_d \left( \frac{\mu}{T} \right)$ appearing in eq. (\ref{AreaforddimSQhightemp}) is given as follows
	\begin{equation}
	\gamma_d\left(\frac{\mu}{T}\right)=N(1)+\frac{d^2(d-2)}{16\pi^2(d-1)}
	\left(\frac{\mu}{T}\right)^2\int_{0}^{1}dx\left(\frac{x\sqrt{1-x^{2(d-1)}}}{\sqrt{1-x^d}}\right)
	\left(\frac{1-x^{d-2}}{1-x^d}\right)+\mathcal{O}\left(\frac{\mu}{T}\right)^4,\label{gammad}
	\end{equation}
	where the constant $N(\varepsilon)$ is calculated at $\varepsilon=1$ and is given as
	\begin{equation}
	N(\varepsilon)=2\left[\frac{\sqrt{\pi } \Gamma \left(-\frac{d-2}{2 (d-1)}\right)}{2 (d-1) \Gamma \left(\frac{1}{2
			(d-1)}\right)}\right]+2\int_{0}^{1}dx
	\left(\frac{\sqrt{1-x^{2(d-1)}}}{x^{d-1}\sqrt{f(z_H x)}}-\frac{1}{x^{d-1}\sqrt{1-x^{2(d-1)}}}\right).\label{Nepsilon}
	\end{equation}
	
	\subsection{Non-extremal RN-$\mathrm{AdS_{d+1}}$ black holes (Large chemical potential - low temperature)}\label{B3}
	
	The dimension dependent constants $ N_{0},N_{1}$ appearing in eq. (\ref{AreaforddimLQhightemp}) are given as follows
	\begin{align}
	N_0&=2\left[\frac{\sqrt{\pi } \Gamma \left(-\frac{d-2}{2 (d-1)}\right)}{2 (d-1) \Gamma \left(\frac{1}{2
			(d-1)}\right)}\right]+2\int_{0}^{1}dx\left(\frac{\sqrt{1-x^{2(d-1)}}}{x^{d-1}
		\sqrt{1-b_0 x^d+b_1 x^{2(d-1)}}}-\frac{1}{x^{d-1}\sqrt{1-x^{2(d-1)}}}\right),\label{N0} \\
	N_1&=\int_{0}^{1}dx\left(\frac{x\sqrt{1-x^{2(d-1)}}}{\sqrt{1-b_0 x^d+b_1 x^{2(d-1)}}}\right)
	\left(\frac{1-x^{d-2}}{1-b_0 x^d+b_1 x^{2(d-1)}}\right).\label{N1}
	\end{align}

\end{appendices}

\bibliographystyle{utphys}

\bibliography{reference}

\end{document}